\documentclass{aa}
\usepackage{showyourwork}
\usepackage[version=4]{mhchem}

\usepackage{txfonts}
\usepackage{natbib}

\usepackage{graphicx}
\usepackage{booktabs}

\usepackage{color}
\usepackage{hyperref}
\hypersetup{colorlinks=true,allcolors=[rgb]{0,0,0.8}}

\makeatletter
\renewcommand*\aa@pageof{, page \thepage{} of \pageref*{LastPage}}
\makeatother

\usepackage[nolist,nohyperlinks]{acronym}

\begin{acronym}
    \acro{ao}[AO]       {adaptive optics}
    \acro{eris}[ERIS]   {Enhanced Resolution Imager and Spectrograph}
    \acro{psf}[PSF]     {point spread function}
    \acro{app}[APP]     {apodizing phase plate}
    \acro{gvapp}[gvAPP] {grating vector apodizing phase plate}
    \acro{vapp}[vAPP]   {vector apodizing phase plate}
    \acro{sphere}[SPHERE]{Spectro-Polarimetric High-contrast Exoplanet REsearch}
    \acro{vlt}[VLT]     {Very Large Telescope}
    \acro{iwa}[IWA]     {inner working angle}
    \acro{owa}[OWA]     {outer working angle}
    \acro{adi}[ADI]     {angular differential imaging}
    \acro{rdi}[RDI]     {reference differential imaging}
    \acro{lcp}[LCP]     {liquid crystal polymer}
    \acro{pca}[PCA]     {principal component analysis}
    \acro{dit}[DIT]     {detector integration time}
    \acro{sed}[SED]     {spectral energy distribution}
    \acro{nix}[NIX]     {Near Infrared Camera System}
    \acro{lcc}[LCC]     {Lower Centaurus Crux}
    \acro{scocen}[Sco-Cen]{Scorpius-Centaurus}
    \acro{trap}[TRAP]   {temporal reference analysis of planets}
    \acro{vosa}[VOSA]   {Virtual Observatory SED Analyzer}
    \acro{irdap}[IRDAP] {IRDIS Data reduction for Accurate Polarimetry}
    \acro{elt}[ELT]     {Extremely Large Telescope}
    \acro{ee}[EE]     {encircled energy}
\end{acronym}

\begin{document} 

\authorrunning{Kenworthy et al.}
\titlerunning{The VLT/ERIS gvAPP coronagraph design and on-sky performance}
  \title{The VLT/ERIS grating vector Apodizing Phase Plate coronagraph} 
  \subtitle{Design and on-sky performance}

\author{M. A. Kenworthy\inst{1}
\and
F. A. Dannert\inst{2}
\and
J. Hayoz\inst{2}
\and
D. Doelman\inst{1,3}
\and
B. J. Sutlieff\inst{4}
\and
P. Liu\inst{1,4}
\and
F. Snik\inst{1}
\and
M. J. Bonse\inst{5,6}
\and
S. P. Quanz\inst{2,7}
\and
C. U. Keller\inst{8,1}
\and
O. Absil\inst{9}\fnmsep\thanks{F.R.S.-FNRS Research Director}
\and
G. Orban de Xivry\inst{9}
\and
R. J. De Rosa\inst{10}
\and
C. Ginski\inst{11}
\and
X. Chen\inst{4}
\and
A. Zurlo\inst{12,13}
\and
B. A. Biller\inst{4}
\and
J. L. Birkby\inst{14}
\and
A. Baruffolo\inst{15}
\and
Y. Dalliliar\inst{16}
\and
R. Davies\inst{17}
\and
M. Dolci\inst{18}
\and
H. Feuchtgruber\inst{17}
\and
A. Glauser\inst{2}
\and
P. Grani\inst{19}
\and
K. Kravchenko\inst{17}
\and
M. MacIntosh\inst{20}
\and
A. Puglisi\inst{19}
\and
C. Rau\inst{17}
\and
A. Riccardi\inst{19}
\and
E. Sturm\inst{17}
\and
W. Taylor\inst{20}
}

\institute{Leiden Observatory, Leiden University, Postbus 9513, 2300 RA Leiden, The Netherlands
 \and
ETH Zurich, Institute for Particle Physics \& Astrophysics, Wolfgang-Pauli-Str. 27, 8093 Zurich, Switzerland
 \and
SRON Netherlands Institute for Space Research, Niels Bohrweg 4, 2333 CA, Leiden, The Netherlands
 \and
Institute for Astronomy, University of Edinburgh, Royal Observatory, Blackford Hill, Edinburgh EH9 3HJ, UK
 \and
European Southern Observatory, Karl-Schwarzschild-Stra{\ss}e 2, 85748 Garching bei M\"unchen
 \and
Max Planck Institute for Intelligent Systems, Max-Planck-Ring 4, 72076 T\"ubingen, Germany
 \and
ETH Zurich, Department of Earth and Planetary Sciences, Sonneggstrasse 5, 8092 Zurich, Switzerland
 \and
National Solar Observatory, 3665 Discovery Dr., Bounder, Co 80303, USA
 \and
STAR Institute, Universit\'e de Li\'ege, All\'ee du Six Ao\^ut 19c, B-4000 Li\'ege, Belgium
 \and
European Southern Observatory, Alonso de C\'{o}rdova 3107, Vitacura, Casilla 19001, Santiago, Chile
 \and
School of Natural Sciences, Center for Astronomy, University of Galway, Galway, H91 CF50, Ireland
 \and
Instituto de Estudios Astrof\'isicos, Facultad de Ingenier\'ia y Ciencias, Universidad Diego Portales, Av. Ej\'ercito Libertador 441, Santiago, Chile
 \and
Millennium Nucleus on Young Exoplanets and their Moons (YEMS)
 \and
Astrophysics, University of Oxford, Denys Wilkinson Building, Keble Road, Oxford, OX1 3RH, United Kingdom
 \and
INAF - Osservatorio Astronomico di Padova, Vicolo dell'Osservatorio 5, 35122, Padova, Italy.
 \and
Physikalisches Institut, Universit{\"a}t zu K{\"o}ln, Z{\"u}lpicher Str. 77, 50937, K{\"o}ln, Germany
 \and
Max-Planck-Institut f{\"u}r extraterrestrische Physik, Postfach 1312, 85741, Garching, Germany
 \and
INAF - Osservatorio Astronomico d'Abruzzo, Via Mentore Maggini, 64100, Teramo, Italy
 \and
INAF - Osservatorio Astrofisico di Arcetri, largo E. Fermi 5, 50125 Firenze, Italy
 \and
STFC UK ATC, Royal Observatory Edinburgh, Blackford Hill. Edinburgh EH9 3HJ, UK
 }

   \date{Received January 26, 2026; accepted March 6, 2026}

  \abstract
   {}
   {We describe the design, laboratory manufacture, and on-sky testing of the \ac{gvapp} coronagraph for the Enhanced Resolution Imager and Spectrograph (ERIS) on the Very Large Telescope.}
   {We used both laboratory measurements and on-sky observations to characterise the \ac{gvapp} in several different filters, from the $K$ to the $L$ band.}
   {In testing, the \ac{gvapp} reaches its design specification in the transmission of the optic with 90\% in the $K$ bands and 60\% in the $L$ band.
   While the \ac{gvapp} reaches its designed raw contrast performance of $1\times10^{-5}$, it does not reach the post-processed contrast of $5\times10^{-5}$ in on-sky observations.
   Electronic detector noise, due to the Airy core of the coronagraphic point spread function inducing cross-talk between the readout amplifiers, produces a repeated pattern within the coronagraphic regions of the \ac{gvapp}.}
   {Despite these limitations, we recommend the gvAPP as a tool for characterising substellar companions with known separations and position angles, which allow them to be placed in the coronagraphic dark holes for observations.
   The \acs{eris} \ac{gvapp}'s leakage term can also be used as a photometric reference for time series observations; however, we caution that the contrast performance may limit such studies to only the brightest targets.
   \acs{eris} \ac{gvapp} data quality may be improved further with better modelling of detector electronic noise.
   This work is a pathfinder for Extremely Large Telescope instruments including METIS, which will include \ac{gvapp} coronagraphs with improved designs based on these results.}


   \maketitle
%

\section{Introduction}

The \acl{eris} \citep[\acs{eris};][]{2023A&A...674A.207D} is a diffraction-limited imager and spectrograph for the Very Large Telescope (VLT).%
\acused{eris}
It is composed of two parts, the integral field spectrograph SPIFFIER \citep{George16} and a diffraction-limited 1 to 5 micron high-angular-resolution imager, the \acl{nix} \citep[\acs{nix};][]{2016SPIE.9908E..3FP}, which has two pixel scales (13 mas covering $26.4''\times26.4''$ and 27 mas covering $55.4''\times55.4''$).
\acused{nix}The science goals of \ac{eris} include the detection and characterisation of extrasolar planets and circumstellar disks around bright stars \citep[e.g.][]{2025A&A...698A..52M}.

The diffraction halo of the central star overwhelms the signals from objects adjacent to the star, making studies of these sources difficult to impossible.
Coronagraphs are angular filters -- they suppress the wavefront from point-like sources on the sky whilst simultaneously allowing wavefronts from adjacent astrophysical sources to pass through to the science camera focal plane  (see \citealt{2025ARA&A..63..179K} for a review).
Focal-plane coronagraphs use a combination of amplitude and phase modifications so that the propagated starlight in the subsequent pupil plane can be blocked by an appropriately matched mask.
Pupil-plane coronagraphs achieve this goal by modifying the wavefront in the pupil plane to redistribute the diffraction structure of the \ac{psf} in the science camera focal plane.
At infrared wavelengths, these coronagraphs have the advantage that the stellar suppression is independent of chopping and telescope pointing. 
In addition, the flux of the rearranged starlight makes image alignment easier in post-processing.
The \ac{eris} \ac{nix} instrument is equipped with two coronagraphs, a grating vector-apodized pupil
plane coronagraph \citep{otten2017sky} and a vector vortex coronagraph \citep{2024SPIE13097E..15O}.

\begin{figure*}
    \centering
    \script{plot_image_phase_psf.py}
    \includegraphics[width=0.95\textwidth]{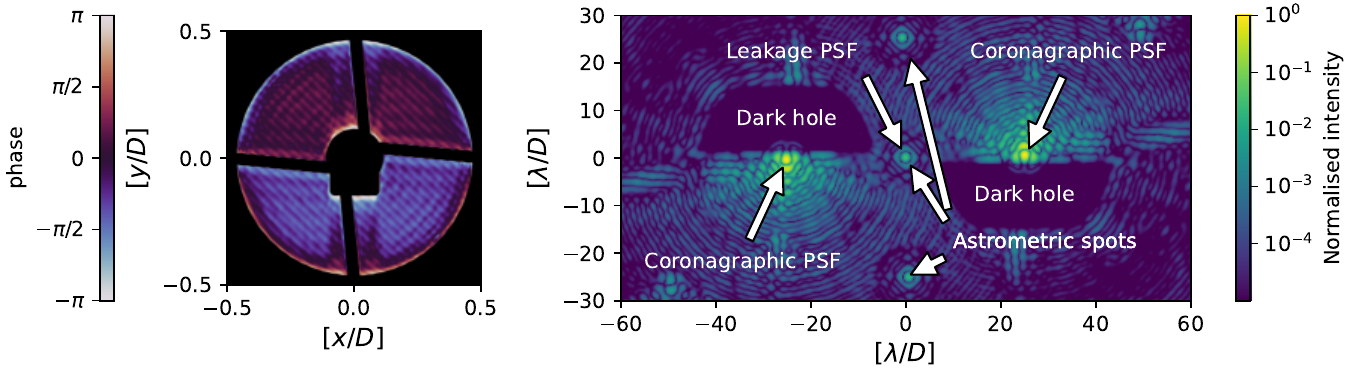}
    \caption{ \ac{eris} \ac{gvapp} phase pattern design and simulated \ac{psf}.
    Left: Phase within the defined \ac{gvapp} pupil.
    Right: Resultant three \acp{psf} of the \ac{gvapp}.\ We show the two coronagraphic \acp{psf} and the central leakage term.
    Three astrometric reference spots can be seen.
    The scale is logarithmic in normalised intensity.}
    \label{fig:gvAPP_phase}
\end{figure*}

The \acl{app} \citep[\acs{app};][]{Codona04} adds a specific phase pattern in the telescope pupil without changing the amplitude of the electric field.
This phase pattern modifies the \ac{psf} of the instrument and suppresses diffraction structures in a 180-degree-wide semi-circular region next to a target star, referred to as a `dark hole' (see Fig.~\ref{fig:gvAPP_phase}).
\acused{app}%
This is demonstrated on-sky in \citet{Kenworthy07} using the variation in thickness of a \ce{ZnSe} substrate to induce the designed phase pattern at one central wavelength using an \ac{app} coronagraph \citep{Kenworthy10a} with NAOS/CONICA \citep{Lenzen03,Rousset03}; this resulted in the characterisation of Beta Pictoris b \citep{2010ApJ...722L..49Q} and the discovery of the substellar companion HD~100546b \citep{Quanz13}.
The original implementation of the \ac{app} does not work for broader bandwidths due to the chromatic nature of available substrates.
An alternative approach provides the ability to make phase-based coronagraphs achromatic over much wider bandwidths: by using the geometric phase \citep[the Pancharatnam-Berry phase;][]{Pancharatnam,Berry}, almost any phase pattern can be encoded in the orientation of the fast axis of an optically active polymer.

Unpolarised light can be decomposed into equal amounts of left and right circularly polarised light, and the vast majority of stars and planets emit almost completely unpolarised light.
Using the geometric phase, both circular polarisations will have the phase pattern encoded in them, but with opposite signs of the given phase pattern.
The 180$^\circ$ dark hole phase pattern is anti-symmetric, which means that the dark hole will be symmetrically point-flipped to the opposite side of the star, thereby providing full dark hole azimuthal coverage around the star.
A quarter-wave plate and Wollaston prism placed after the optically active polymer then splits the two opposite circularly polarised beams towards different locations in the focal plane \citep{Otten14}, separating the two coronagraphic \acp{psf} (e.g. Fig.~\ref{fig:gvAPP_phase}, right panel).
A single layer of a given polymer is highly chromatic, but combining several carefully specified layers with different chromatic properties, each of which self-align with the layer beneath it, allows achromaticity to be generated over an octave in bandwidth, resulting in a \acl{vapp} \citep[\acs{vapp};][]{Snik12}.
 The addition of a sinusoidally periodic phase ramp into the phase pattern generates a polarisation diffraction grating in the pupil beam.
This diffracts the two circular polarisations into the $m=\pm1$ orders, separating them at the focal plane and removing the need for the Wollaston prism and quarter-wave plate: this is the principle behind the \acl{gvapp} \citep[\acs{gvapp};][]{otten2017sky}.
\acused{gvapp}There are now \ac{gvapp} coronagraphs installed at telescopes worldwide, enabling a range of scientific goals: searches for directly imaged exoplanets \citep{2018SPIE10703E..2TL, 2023AJ....165..216L, 2025A&A...693A..81B, 2025MNRAS.536.1455M}, the spectroscopic characterisation of known substellar companions, including the HR~8799 exoplanets \citep{2021MNRAS.506.3224S, 2022AJ....163..217D}, imaging of the protoplanetary disk around PDS~201 \citep{2020AJ....159..252W}, variability monitoring of the substellar companion HD~1160~B \citep{2023MNRAS.520.4235S, 2024MNRAS.531.2168S}, and the development of focal-plane wavefront sensing and spatial linear dark field control \citep{2019A&A...632A..48B, 2019JATIS...5d9004M, 2021A&A...646A.145M}.
An overview of existing and upcoming \ac{gvapp} coronagraphs is found in \citet{2021ApOpt..60D..52D}.

In this paper we report on the theoretical, laboratory, and on-sky performance of the \ac{gvapp} coronagraph in the \ac{eris} \ac{nix} instrument.
We discuss the phase pattern design, construction, and cryogenic testing in Sect. \ref{sec:phase_design}.
We then describe the on-sky measurements, including \ac{psf} validation and the attained contrast curves, in Sect. \ref{sec:on-sky}, before analysing the performance of the \ac{gvapp} and making recommendations for its use in Sect.~\ref{sec:performance_analysis}.
Finally, we highlight our conclusions in Sect. \ref{sec:conclusion}.

\section{Design and construction}
\label{sec:phase_design}

\subsection{Optical design}

We used a \ac{gvapp} design algorithm that can find dark hole solutions for arbitrarily shaped telescope pupils and dark hole geometries \citep{2021ApOpt..60D..52D}.
We took the \ac{eris} pupil geometry to be the VLT primary mirror, secondary support structure, and secondary mirror, with a rectangular extension next to the secondary mirror caused by the tertiary mirror raised up into its stow position (see Fig.~\ref{fig:gvAPP_phase}, left panel).
A thin metal mask (the `amplitude mask') in the \ac{gvapp} both defines the pupil for the coronagraph and acts as a cold stop, preventing thermal emission from the telescope support structures reaching the science camera focal plane.

In order for the \ac{gvapp} pupil to be fully illuminated by the  telescope pupil within \ac{eris} (which is subject to flexure and alignment tolerances), the \ac{gvapp} optic is undersized ($d=11.50$\,mm) with respect to the nominal diameter of the reimaged \ac{eris} pupil ($d=11.76\,$mm).
The design parameters are listed in Table~\ref{tab:design_param}.

In monochromatic light, the \ac{gvapp} produces three \acp{psf}.
Two of these are coronagraphic \acp{psf} with a dark hole on one side of the \ac{psf} Airy core of the target star, matching the \ac{psf} of the \ac{gvapp} pupil phase pattern (see Fig.~\ref{fig:gvAPP_phase}).
These two \acp{psf} are displaced by equal distances on either side of a central leakage term \ac{psf}, which corresponds to the \ac{psf} of the amplitude mask alone (i.e. an unmodified \ac{psf} of the target star at a fraction of its original flux).
The majority of the stellar flux is divided equally between the two coronagraphic \acp{psf}, and a small fraction remains in the leakage term \ac{psf}.
The \ac{eris} \ac{gvapp} design has two additional faint \acp{psf} that are added perpendicular to the coronagraphic \ac{psf} separation axis -- these act as astrometric references to align the images if the cores of the coronagraphic \acp{psf} are over-saturated.
A diffraction grating pattern is added to the vAPP phase pattern: this diffraction grating of 25 cycles across the \ac{gvapp} pupil means that the separation between the coronagraphic \acp{psf} are separated 25$\lambda/D$ from the leakage \ac{psf}, and a similar effect occurs for the astrometric reference \acp{psf}.
Since the location of the coronagraphic \acp{psf} (and any associated faint companion next to the star) scales with wavelength, any faint companions that are in one of the dark holes will have smearing along a line between the companion and leakage \ac{psf}.
Thus, although the \ac{gvapp} can work at all wavelengths accessible to ERIS, to avoid excessive smearing, the \ac{gvapp} should only be used with narrow band filters with a maximum fractional bandwidth of 6\% \citep{2022MNRAS.515.5629D}.

 \begin{table}
 \caption{Design parameters for the \ac{eris} \ac{gvapp} including the \ac{iwa} and \ac{owa}.         \label{tab:design_param}
}
     $$ 
         \begin{array}{p{0.5\linewidth}l}
            \hline
            \noalign{\smallskip}
            Parameter     &  Value  \\
            \noalign{\smallskip}
            \hline
            \noalign{\smallskip}
            IWA $[\lambda/D]$             &  2.2    \\
            IWA contrast                  & 10^{-4} \\
            OWA $[\lambda/D]$             &  15     \\
            OWA contrast                  & 10^{-5} \\
            Encircled energy of Airy core & 50.9\%  \\
            \noalign{\smallskip}
            \hline
         \end{array}
     $$ 
\end{table}
   
Two \acp{gvapp} have been manufactured by ImagineOptix in 2018, and the best \ac{gvapp} chosen based on the transmission in the operating wavelength range.
The manufacturing method of the patterned liquid-crystal layers is identical to the MagAO \ac{gvapp} \citep{otten2017sky} and is described in detail in \citet{kim2015fabrication}. 
Here, we outline the optical assembly unique to the \ac{eris} \ac{gvapp}. 

The \ac{gvapp} consists of four individual \ce{CaF2} substrates glued together.
Four substrates are required due to two manufacturing limitations.
First, the anti-reflection (AR) coating cannot be applied to the opposite side of the substrate of the amplitude mask, resulting in the need for a separate AR-coated substrate of 1 mm thickness. 
Second, it is not possible to apply the liquid-crystal film to the wedged substrate because of the wedge angle of 0.44 degrees.
The four substrates have a total thickness of 7.55 mm at the thick end of the wedge (7.39 mm at the thin end of the wedge) with an overall measured diameter of 20.95 mm.
The substrates are glued together with NOA-61 and aged 10 hrs at $50^\circ$C.
The slope of the wedge is perpendicular to the line of the three PSFs and chosen to prevent reflection ghosts from the coronagraphic \acp{psf} appearing in the \ac{gvapp} coronagraphic holes.
The amplitude mask is made of aluminium with a thickness $\sim 300$ nm.
We confirmed that the centre of the clear aperture of the mask is 0.020 mm from the centre of the 21 mm substrates.
The AR coatings have $< 0.5$ \% average reflectance each.
ImagineOptix report an alignment error of $<1$ micron between the centre of the liquid-crystal pattern and the centre of the aluminium mask, although this was not independently verified.

\subsection{Optical throughput of the gvAPP}\label{sec:optical_tput}
\begin{figure}
    \centering
    \script{plot_throughput.py} 
    \includegraphics{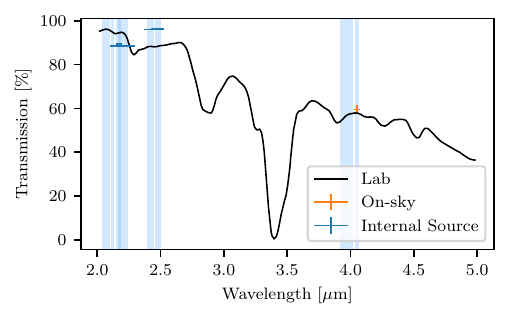}
    \caption{Optical transmission of the \ac{gvapp} measured in the laboratory (black curve) and the on-sky background transmission in the Br-$\alpha$ filter (orange point).
    Additional measurements with the \ac{eris} internal calibration unit are shown by the blue points. 
    The blue regions trace all filters available for observations with the \ac{gvapp} in \ac{eris}.
    Note that for on-sky throughput, this transmission curve should be multiplied by 0.23 to account for the EE and division of the flux between coronagraphic \acp{psf} (see Sect.~\ref{sect:eris_gvapp_throughput}).
    }
    \label{fig:gvapp_transmission}
\end{figure}

The \ac{eris} \ac{gvapp} transmission (measured by ImagineOptix) is shown in Fig.~\ref{fig:gvapp_transmission} and is measured in a clear aperture portion of the \ac{gvapp} using a Thorlabs OSA205 Fourier spectrometer. 
The measurement includes all substrates, glue layers, \ac{lcp} layer, and AR coatings.
The overall transmission in is $\sim 90\%$ in the $K$ band, $\sim 50\%$ in the $L$ band, and $\sim 40\%$ in the $M$ band.
The transmission curve is comparable to the Large Binocular Telescope (LBT) double-grating vAPP (dgvAPP) transmission curve found in \cite{2022AJ....163..217D}.
Like the LBT dgvAPP, the absorption feature at $3.3 \mu$m goes to 0\% transmission, rendering $L$-short observations ineffective. 
This absorption feature is caused by absorption in both the glue and liquid-crystal layers due to vibrational modes of chemical bonds with carbon atoms \citep{otten2017sky}.

To verify the transmission in the $K$ band, we used the \ac{eris} internal calibration unit to image the \acp{psf} both with and without the \ac{gvapp} in Br-$\gamma$, $K$-peak, IB-2.42, and IB-2.48 filters at two different positions on the detector\footnote{See the ERIS instrument description for filter definitions: \url{https://www.eso.org/sci/facilities/paranal/instruments/eris/inst.html}.}. 
We subtracted the background and used aperture photometry on the \ac{psf} cores to obtain the fluxes of the \acp{psf}, including both coronagraphic \acp{psf} and the leakage \ac{psf}.
The \ac{ee} of the Airy core of the coronagraphic \ac{app} \acp{psf} is reduced due to the nature of the pupil apodisation. 
The \ac{app} design is a trade-off between coronagraphic suppression and \ac{ee} fraction.  
This is calculated by simulating the non-coronagraphic \ac{psf} in addition to the \ac{app} \ac{psf} using {\tt  HCIPy} \citep{por2018hcipy} and performing aperture photometry enclosing the central Airy core to the first minimum. 
We calculated the correction for the undersized pupil using archival pupil images and comparing the area of these pupils on the detector.
The phase apodisation reduces the \ac{ee} transmission to $47.6\%$ and the undersizing of the pupil has a relative transmission of $80\%$ compared to no pupil mask. 
We divided out these factors to calculate the transmission of the \ac{app} optic from the fluxes from the aperture photometry, as shown in Fig.~\ref{fig:gvapp_transmission}.

\section{On-sky measurements}\label{sec:on-sky}

\begin{table*}
    \centering
         \caption{Observing log.}
    \renewcommand*{\arraystretch}{1.2}
     \begin{tabular}{l l l c c c c c c c c c}
        \hline
        UT (mid) & Target & Filter & $\lambda_{\mathrm{c}}$ & $\Delta \lambda$ & $m_{Filter}$ & $t_{\mathrm{DIT}}$ & $t_{\mathrm{total}}$ & $X$ & Seeing & $\tau$ & $\Delta\theta$ \\
         & ~ & ~ & [$\mu$m] & [$\mu$m] & [mag] & [s] & [min] & ~ & [$\arcsec$] & [ms] &  [$^{\circ}$] \\
        \hline
        2022-07-09T08:24 & $\gamma$~Gru & Br-$\gamma$      & 2.172 & 0.020 & 3.45 & 0.3 & 48  & 1.05 & 0.88 & 2.8  & 97 \\
        2022-11-05T01:47 & HR~8799      & Br-$\alpha$-cont & 3.964 & 0.104 & 5.23 & 4.0 & 53  & 1.49 & 0.71 & 5.0  & 117 \\
        2022-07-07T05:26 & $\iota$~Cap  & Br-$\alpha$      & 4.051 & 0.025 & 2.18 & 3.0 & 15  & 1.1  & 0.63 & 4.34 & 11 \\
        2023-10-16T01:47& HR~8799      & $K$-peak           & 2.198 & 0.098 & 5.24 & 0.25 & 189 & 1.49 & 0.83 & 6    & 65 \\
        \hline
     \end{tabular}
    
\tablefoot{The columns $\lambda_{\mathrm{c}}$ and $\Delta \lambda$ denote the central wavelength and effective width of the photometric filters.
         Airmass $X$, seeing, and coherence time $\tau$ are the median values for a given dataset.
         The amount of sky rotation during the exposures is given by $\Delta\theta$.
         The Br-$\gamma$, Br-$\alpha$-cont, and Br-$\alpha$ datasets were obtained during instrument commissioning runs, and the $K$-peak observations collected as part of an open time programme with ESO programme no. 112.25QA.001 (PI: Sutlieff).
         For a single data cube there are $N_{DIT}$ frames each of $t_{DIT}$ integration. 
         Multiple cubes then make up the total integration time $t_{total}$.}\label{tab:ppc_obs_log}
\end{table*}

The on-sky performance of the \ac{eris} \ac{gvapp} in the narrow $K$- and $L$-band filters was evaluated in 2022 and 2023; see Table~\ref{tab:ppc_obs_log} for observing details.
The Br-$\gamma$, Br-$\alpha$-cont, and Br-$\alpha$ filters were tested as part of the instrument commissioning, whereas $K$-peak observations were obtained as part of an open time programme with ESO programme 112.25QA.001 (PI: Sutlieff).
We report on the high-contrast imaging performance using the metrics suggested in the Optimal Optical Coronagraph Workshop \citep{2018SPIE10698E..2SR}, namely the off-axis throughput, the raw contrast, and the post-processed detection limits in the companion-to-star flux ratio.

\subsection{Pre-processing}

\label{sect:preprocessing}

The features of the \ac{gvapp} \ac{psf} -- in particular the lack of rotational symmetry -- render the pre-processing challenging.
The following describes the data processing pipeline that is used to convert the raw detector integrations to science-ready images. 
We used the Python package \texttt{pynpoint} \citep{Amara2012PYNPOINTexoplanets,Stolker2019PynPointdata} to reduce all datasets except for the $K$-peak dataset, which is described in Sect.~\ref{processing_Kpeak}.

\begin{figure}
    \centering
    \script{plot_combine_science.py}
    \includegraphics[width=1.0\columnwidth]{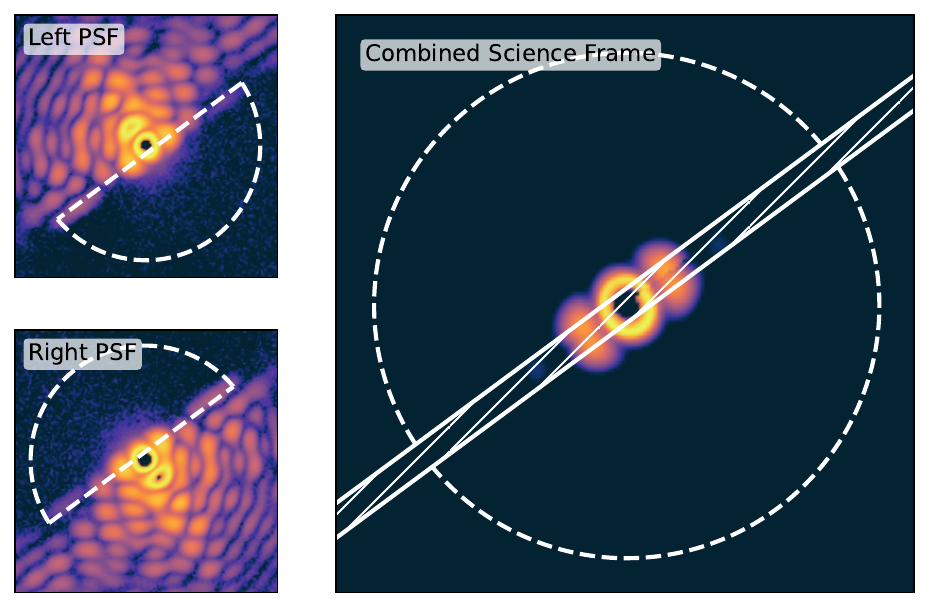}
    \caption{Geometry of combining the two coronagraphic \acp{psf} to form the coverage around the target star.
    The hatched area is masked out during subsequent image combination and processing.
    The semi-circular dashed lines are a visual aid only; the area at wider separations beyond these circles is not masked in the combined image.
    }
    \label{fig:combine}
\end{figure}

The first part of the reduction is similar to any adaptive optics (AO) observation taken without a coronagraph.
To correct the cosmetics of the detector, the science frames are dark and flat corrected, and then bad pixels are identified via sigma-clipping on the dark and flat frames (as well as in the raw science images) and are replaced by the median of the neighbouring pixels.
All our observations follow an `AB' nodding sequence with a telescope offset every 3-5~min in order to subtract the near-IR sky and telescope background.
We refer to the two positions on the detector as Nod A and Nod B.
We subtracted the sky background at a given nodding position by taking the average of the frames at the opposite nodding position before and after the current position.
We subtracted the median of every row and column of the image to remove the remaining detector artefacts presenting as stripes in the image.
The science frames were then aligned using cross-correlation, centred using a Gaussian fit to the leakage term \ac{psf}, and cropped to include the whole \ac{gvapp} \ac{psf} pattern.
For fainter targets, this alignment was done on the coronagraphic \acp{psf}, as their cores are not saturated.
Bad frames during which the AO loop was open were identified by computing the mean-square error of the leakage term \ac{psf} relative to the mean science image.
%
The second part of the reduction is specific to \ac{gvapp} observations and aims at cutting and stitching together the dark holes of the two coronagraphic \acp{psf}.
Since the centres of the coronagraphic \acp{psf} are saturated and are therefore challenging to precisely determine, we identified their position relative to the three astrometric \acp{psf} on the simulated \ac{psf} (see Fig.~\ref{fig:gvAPP_phase}).
We measured the position of the three central astrometric spots in the science frames and inferred the position of the centres of the coronagraphic \acp{psf}.
We then cut out a D-shaped dark hole from each coronagraphic \ac{psf} and put them into a single frame.
The straight edge of the dark hole lies along the transition between the bright and dark part of the \ac{psf} and is offset by 0.5~$\lambda/D$, and so in the combined frame we are missing a 1~$\lambda/D$ wide strip, resulting in incomplete coverage around the central star (see Fig.~\ref{fig:combine}).
This pre-processing pipeline delivers science frames containing the two dark holes of the coronagraphic \acp{psf} stitched together, with the 1~$\lambda/D$ strip parallel to the long side of the `D' masked out.

\subsection{Throughput}\label{sect:eris_gvapp_throughput}

Contrary to focal plane coronagraphs, to first order, the throughput does not depend on the position of the source in the field of view, as the \ac{gvapp} only modifies the wavefront in the pupil plane.
This means that the off-axis throughput affecting a companion can be approximated by the transmission of the \ac{gvapp} element at the respective wavelength.
As described in Sect.~\ref{sec:phase_design}, the transmission of the \ac{gvapp} is measured using a fibre-coupled IR light source and Fourier spectrometer (see Fig.~\ref{fig:gvapp_transmission}).
While the $K$-band transmission of the \ac{gvapp} can be measured using the internal calibration unit (see Sect.~\ref{sec:optical_tput}), its wavelength limitations mean that the $L$- and $M$-band transmission can only be measured on-sky.
A differential infrared background measurement is especially suitable to derive the transmission.
Two consecutive observations are performed with identical \acp{dit} using the Br-$\alpha$ filter, one with the \ac{gvapp} mask in the aperture and one with a mask blocking only the outer telescope support structure.
After dark-subtracting the data, we measured the total counts in a region on the detector with no sources and only a few bad pixels.
Accounting for the different outer aperture sizes of the \ac{gvapp} (12~mm) and the $LM$-pupil mask in ERIS (11.2~mm), the ratio of the total counts yields a transmission of $(59.5 \pm 2.2)\, \%$ in the Br-$\alpha$ filter.
While this on-sky measurement of the transmission agrees with the laboratory measurements, we note that it is slightly conservative as it does not account for the additional thermal contribution from the unobstructed spiders when observing with the $LM$-pupil mask.

The point source throughput with the \ac{gvapp} is multiplied by a  factor of 0.45 to account for the division between the two coronagraphic PSFs and leakage term, and multiplied by an additional factor of 0.51 for the EE of the Airy cores.
This results in a total throughput of $13.6\, \%$ in the Br-$\alpha$ filter and a throughput of $21.8\, \%$ for the K-peak filter.

\subsection{Raw contrast}

\begin{figure*}
    \centering
    \script{plot_raw_contrast_bracont_v2.py}
    \includegraphics[]{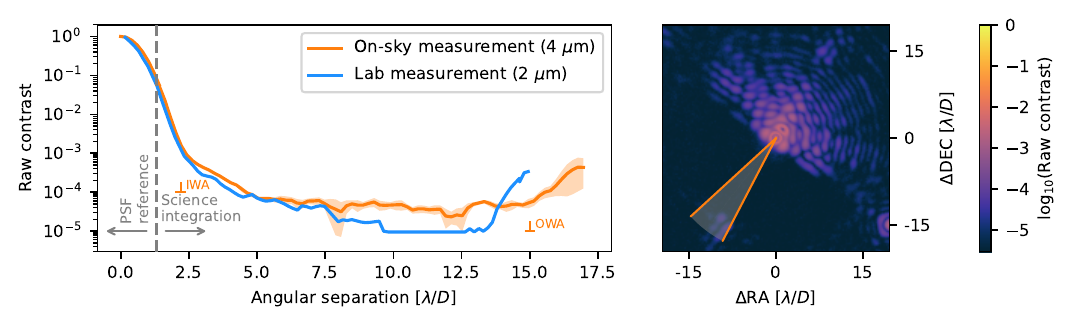}
    \caption{Raw contrast of the \ac{gvapp} in the Br-$\alpha$-continuum filter.
    Left: Raw flux contrast evaluated in $\lambda/D$-sized apertures placed in the dark hole over an aperture placed on the \ac{psf} centre in the left lobe of a deep integration on HR~8799 at $3.96\, \mathrm{\mu m}$.
    The orange line traces the median contrast evaluated in a $\pm 10^\circ$ region perpendicular to the main axis of the \ac{gvapp} \ac{psf}.
    The orange markers indicate the \ac{iwa} and \ac{owa} provided in Table~\ref{tab:design_param}.
    The blue curve displays the raw contrast measured on the \ac{gvapp} element under laboratory conditions at $2 \, \mathrm{\mu m}$ by \citet{2021JATIS...7d5001B} -- the values from 10 to 12.5~$\lambda/D$ are below the sensitivity of the test bench.
    The rise seen at 13.5~$\lambda/D$ in the laboratory data is due to limitations from internal reflections in the test bench not present in ERIS.
    Right: Focal plane image of the left lobe of the \ac{gvapp} on which the raw contrast was evaluated.
    The shaded orange wedge is the $\pm 10^\circ$ region in which the orange contrast curve was measured.
    }
    \label{fig:raw_contrast}
\end{figure*}

The raw contrast is evaluated on the HR~8799 dataset in the Br-$\alpha$-cont filter (see Table~\ref{tab:ppc_obs_log}), which is pre-processed as described in Sect.~\ref{sect:preprocessing}.
To derive the raw contrast as described in \citet{2018SPIE10698E..2SR}, this dataset contains the host star as a single bright point source, which produces a high-S/N image of the \ac{gvapp} coronagraphic \ac{psf}, $\mathrm{PSF_{coro}}$.
This \ac{psf} is sampled in apertures $\mathrm{AP}$ of diameter $\lambda/D$, starting at the coronagraphic \ac{psf} core $\eta_s = \int_{\mathrm{AP}(0)} \mathrm{PSF_{coro}}(x) \, \mathrm{d}x $ and at various places in the dark hole yielding $\eta_p(x_0) = \int_{\mathrm{AP}(x_0)} \mathrm{PSF_{coro}}(x) \, \mathrm{d}x$.
The raw contrast is then reported as $C(x_0) = \eta_p(x_0) / \eta_s$.
As the cores of the \ac{psf} are saturated in the science images, the raw contrast is calibrated using the unsaturated \ac{psf} in the \ac{psf} calibration images or can be calibrated using the leakage \ac{psf} measurements. 
To do so, first the raw contrast is calculated in both the science and in a shorter exposure unsaturated coronagraphic \ac{psf} image. 
The science data raw contrast curve is then scaled to match the \ac{psf} reference curve in the unsaturated and high-S/N region from $1.5 - 2 \, \lambda/D$.
As can be seen in Fig.~\ref{fig:raw_contrast}, the deepest raw contrast is reached when evaluated perpendicularly to the main axis of the coronagraphic \ac{psf}.
In this case, the on-sky raw contrast performance in the $L$ band closely follows the laboratory measurement \citep{2021JATIS...7d5001B} out to about $8 \, \lambda/D$.

The \ac{gvapp} is designed to provide high-contrast in 90\% of the region between $2.2$ and $15\lambda/D$ via its two dark holes.
It is therefore instructive to evaluate the raw contrast throughput along the extent of the dark hole.
This is done by placing circular apertures of diameter 1~$\lambda/D$ out to a maximum opening angle of $81^\circ$ perpendicular to the \ac{psf} main axis (i.e. the axis perpendicular to the edge of the dark hole, passing through the coronagraphic \ac{psf} core).
We find that between 4 and 15~$\lambda/D$, the raw contrast does not significantly change out to $60^\circ$ away from the perpendicular, allowing for a $\sim 70 \, \%$ sky access at full raw contrast performance.
Beyond $60^\circ$ up to the designed maximum angle of $81^\circ$, the raw contrast starts to be impacted by apertures that partially overlap with the bright side of the coronagraphic \ac{psf}.
In these regions, full contrast performance is only available from $10$ to $15 \, \lambda/D$.

The median raw contrast performance at $\pm 10 ^\circ$ to the perpendicular is measured to $1.6 \times 10^{-3}$ at $2.2 \, \lambda/D$ and to an upper limit of $5.3 \times 10^{-5}$ at $15 \, \lambda/D$.
Compared to the design parameters in Table~\ref{tab:design_param}, this means that the \ac{gvapp} underperforms by more than an order of magnitude at the inner working angle (IWA).
An analysis by \citet{2021JATIS...7d5001B} shows that polishing errors (with spatial frequencies following a power law) in the reimaging optics can result in a scattered light halo that is consistent with this reduced performance.
Microscope measurements show that the manufactured LCP pattern is not the cause of this decrease in raw contrast.
At the outer working angle (OWA), the designed contrast performance lies below the sky background limit and cannot be verified on-sky without a longer on-sky integration or a brighter target star.

\subsection{Post-processed contrast: Narrow-band filters}

To characterise the post-processed detection limits of the \ac{eris} \ac{gvapp} in the narrow $K$ and $L$ bands, the stars $\gamma$~Gru ($K=3.45$~mag), HR~8799 ($K=5.24$~mag, $L=5.23$~mag), and $\iota$~Cap ($L=2.18$~mag) were observed using the Br-$\gamma$, Br-$\alpha$, and Br-$\alpha$-continuum filters during commissioning. 
All three observations follow the same sequence in pupil-tracking mode, to profit from \ac{adi} post-processing \citep[][]{2006ApJ...641..556M}.
We first recorded a \ac{psf} reference using a \ac{dit} short enough to avoid any saturation of the \ac{gvapp} \ac{psf}.
For the main observing sequence, the \ac{dit} is increased such that the two main \ac{gvapp} \ac{psf} features are saturated but the central leakage term remains unsaturated.
Every 3~min, the star is nodded by 8\arcsec{} from Nod A to Nod B to estimate and remove the near-IR background.
Every \ac{dit} is written out in cube mode. 
The short observation time in Br-$\alpha$ sequence result in a very small field rotation, and the weather conditions were worse than average, resulting in several poor-quality and open AO loop frames within each data cube, which are removed. 
After pre-processing the data as described in Sect.~\ref{sect:preprocessing}, the data are post-processed using \ac{pca}/\ac{adi} \citep{Amara2012PYNPOINTexoplanets}.
We calculated the detection limits for each filter by performing fake planet injection tests using \texttt{applefy} \citep{Bonse2023ComparingNoise}.
The fake planet templates were constructed from the central leakage term of the unsaturated \ac{psf} reference and were normalised to match the flux of the coronagraphic \ac{psf} using the unsaturated coronagraphic \ac{psf} reference image.
The contrast curves were then created by selecting the principal component resulting in the deepest contrast at each angular separation \citep[i.e. optimal PCA/ADI;][]{Meshkat14} with the bright 1~$\lambda/D$ strip masked out in subsequent calculations.

\subsection{Post-processed contrast: K-peak filter}\label{processing_Kpeak}

The deepest set of observations are 189\,min of data taken in the $K$-peak filter (ESO programme 112.25QA.001, PI: Sutlieff), which has a bandwidth of $0.10 \mu$m.
The median airmass, seeing, and coherence time $\tau$ for the observations obtained with the $K$-peak filter (which was designed by \citealt{2022MNRAS.515.5629D}) are given in Table~\ref{tab:ppc_obs_log}.
The data were reduced by adapting the method applied by \citet{2021MNRAS.506.3224S} to MagAO vAPP data.
The bad pixel correction was carried out using the ESO's ERIS-NIX bad pixel map\footnote{Available at \url{https://www.eso.org/sci/facilities/paranal/instruments/eris/img/master_bpm_lamp.fits.gz}.}.
An averaged dark frame was created by taking the median of five dark frames, which was then subtracted from every frame in the data.
We flat-fielded the data by dividing every frame by a flat frame obtained with the same window size as the data, which we normalised by dividing it by the median value in a region clear of large bad pixel clusters.
Background subtraction was carried out using the frames from the opposite nod position.
To reduce the computational time, every 360 frames were binned, which resulted in 63 frames per nodding position.
We aligned the frames using the {\tt phase\_cross\_correlation} function included in the {\tt scikit-image} \citep{scikit-image} Python package.
We then corrected for row and column systematics by subtracting the median values along each axis, as calculated with the \ac{gvapp} \acp{psf} masked.
We applied the \acl{trap} \citep[\acs{trap};][]{Samland2021TRAP:Separations} model to the binned dataset following the optimised method for \ac{gvapp} coronagraphic datasets demonstrated in \cite{LiuP2023}.
There are four sub-datasets: each nodding position contains left and right complementary coronagraphic images (e.g. see Fig.~\ref{fig:detector_pattern}).
We reduced the four sub-datasets independently by taking the stellar \ac{psf} in each frame as the \ac{psf} model for that frame and selected reference pixels from both the bright and dark sides to build noise models.
We derived a contrast map and an un-normalised uncertainty map for each sub-dataset.
We combined the contrast maps and uncertainty maps weighted by the uncertainty, firstly for the left and right coronagraphic datasets and secondly for the two nodding positions. 
The normalisation factor for the final uncertainty was calculated from the combined contrast map divided by the combined uncertainty map. 
The 5$\sigma$ contrast curve was built by multiplying five times the combined uncertainty with the normalisation factor.

\subsection{Contrast curves}

\begin{figure}
    \centering
    \script{plot_contrast_curves_updated.py}
    \includegraphics[width=1.0\columnwidth]{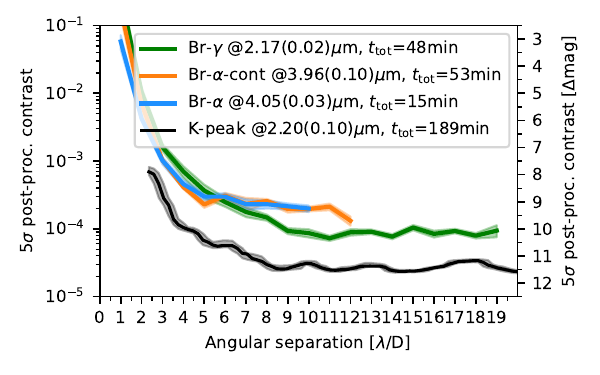}
    \caption{Post-processed contrast of the \ac{eris} \ac{gvapp} in the Br-$\gamma$, Br-$\alpha$, Br-$\alpha$-continuum, and $K$-peak filters.
    The contrast curves for the first three filters were calculated after \ac{psf} subtraction with optimal \ac{pca}/\ac{adi} via fake planet injections.
    The contrast curve for the $K$-peak filter was calculated with the TRAP algorithm \citep{Samland2021TRAP:Separations}.
    The angular separation is shown in units of $\lambda/D$ in order to compare the results for the different narrow band filters.
    The central wavelength and effective width of each filter are indicated in the label of each curve.
    The shaded area represents the uncertainty of the contrast curve.}
    \label{fig:contrast_curves}
\end{figure}

The resulting contrast curves for each narrow-band filter are shown in Fig.~\ref{fig:contrast_curves}.
The three Brackett curves approximately match each other up to 6~$\lambda/D$, where they reach a contrast of $2.5\times 10^{-4}$ (9~magnitudes).
From that point outwards, the two $L$-band contrast curves flatten due to the increased near-IR background from the atmosphere and warm telescope optics.
The Br-$\gamma$ curve flattens at $10^{-4}$ (i.e. 10~mag) around 9-10~$\lambda/D$.
The $K$-peak observation reach a contrast deeper by 1.5~mag due to the six times longer integration time.
Using the contrast at the largest separation and the corresponding stellar magnitudes in the $K$ and $L$ bands, we computed the sensitivity limits scaled to 1 hour (using the square root of the observing time) in each narrow band.
We report the limits for each filter in Table~\ref{tab:ppc_sensitivity}.

\begin{table}
    \centering
        \caption{gvAPP sensitivity limits in narrow-band filters scaled to 1h total on-sky integration time.\label{tab:ppc_sensitivity}}
    \renewcommand*{\arraystretch}{1.2}
     \begin{tabular}{l r}
        \hline
        Filter & Sensitivity [mag] \\
        \hline
        Br-$\gamma$ & 13.72 \\
        Br-$\alpha$-cont & 14.56 \\
        Br-$\alpha$ & 12.18 \\
        $K$-peak & 15.98 \\
        \hline
     \end{tabular}

\end{table}

\section{Performance analysis}\label{sec:performance_analysis}

\subsection{PSF flux imbalances and sensitivity with the position on the detector}\label{sect:imbalances}

We investigated the relative on-sky throughputs of the central astrometric reference spots in the $K$-peak data by measuring the ratio of the flux of each spot to the leakage term.
To do this, we first extracted the aperture photometry of each spot and the leakage term for each frame in the data.
We then divided the photometry of each spot by that of the leakage term, and took the time-averaged mean in each case.
These values are shown in Table~\ref{table:measured_ratios} for two different nod positions, corresponding to opposite sides of the \ac{eris} NIX detector.
The `left-hand' and `right-hand' positions of the spots are defined relative to the detector orientation (see the annotations in Fig.~\ref{fig:detector_pattern}).
The difference between these values is likely due to cross-talk between the readout lines of the sub-arrays in the \ac{eris} detector.
The left-hand spot has a brighter background than the leakage term than the right-hand spot at the A nod position, while it is the opposite at the B nod position.

\begin{table}
\begin{center}
\caption{Ratio of the coronagraphic \acp{psf} to the leakage term for two nod positions (A and B) on the \ac{eris} detector.}
\begin{tabular}{rcc}
\toprule
PSF ratio &A-nod&B-nod\\
\midrule
Left-hand coronagraphic/leakage  & 5.50\% & 4.79\% \\
Right-hand coronagraphic/leakage & 5.32\% & 5.26\% \\
\bottomrule
\end{tabular}

\label{table:measured_ratios}
\end{center}
\end{table}

\begin{table}
\centering
\caption{Flux ratios of the \acp{psf} in the \ac{gvapp}.}
\begin{tabular}{@{}rccc@{}}
\toprule
                 & $\frac{\mathrm{Coronagraphic\,PSF\,flux}}{\mathrm{Leakage\,flux}}$ & $\frac{\mathrm{Astrometric\,spot\,flux}}{\mathrm{Leakage\,flux}}$ & Asymmetry \\ \midrule
$K$-peak           & 3.16            &  0.059                 &   7.0 \%           \\
Br-$\alpha$-cont &     8.68     &    0.16           &  8.1 \%         \\ \bottomrule
\end{tabular}

\tablefoot{ Asymmetry refers to the average difference in flux between the two coronagraphic \acp{psf}.
We note that for the $K$-peak data,  the brighter \acp{psf} differed between Nod A and Nod B.
The value given here is an average.\label{table:measured_asymmetries}}
\end{table}

The two coronagraphic \acp{psf} have a brightness asymmetry with one appearing brighter than the other for a given nod position on the \ac{eris} detector.
We attribute this difference in flux due to the introduction of circular polarisation within the \ac{eris} optical train, which includes several high incidence reflections that can induce this.
In the $K$-peak data, the brighter coronagraphic \acp{psf} alternated between left-hand and right-hand \acp{psf}, indicating that this effect varies with the position of the star image on the detector (consistent with the explanation above).
Average measurements of this asymmetry for the datasets in this paper are given in Table~\ref{table:measured_asymmetries}.
The consequence for observing with the \ac{gvapp} in ERIS is that the two nod positions need to be reduced independently with separate flux calibrations for the leakage and coronagraphic PSF.

\subsection{Reduced broadband sensitivity}\label{sect:reduced_sensitivity}

The $K$-peak observation reaches a sensitivity that is deeper by 2.2~mag compared to the Br-$\gamma$ curve.
If we take the broader filter into account and assume that the sensitivity is proportional to the square root of the transmitted flux through the filters, then the difference in sensitivity is expected to be 0.8~mag between the Br-$\gamma$ and $K$-peak filters -- assuming that the sky background is the same.
Therefore, it seems that the Br-$\gamma$ observation is not limited by the sky background, but rather by an additional component in the dark hole that could be caused by the brighter star; $\gamma$~Gru -- the star observed for the Br-$\gamma$ contrast curve -- is brighter than HR~8799 by 1.8~mag.
The same reasoning can be applied to the Br-$\alpha$ and Br-$\alpha$-cont observations, for which the broader filter reaches a sensitivity that is deeper by 2.4~mag compared to the narrower filter and where the observed star was fainter by 3~mag.

\begin{figure*}
    \centering
    \script{plot_detector_pattern.py}
    \includegraphics[width=0.95\textwidth]{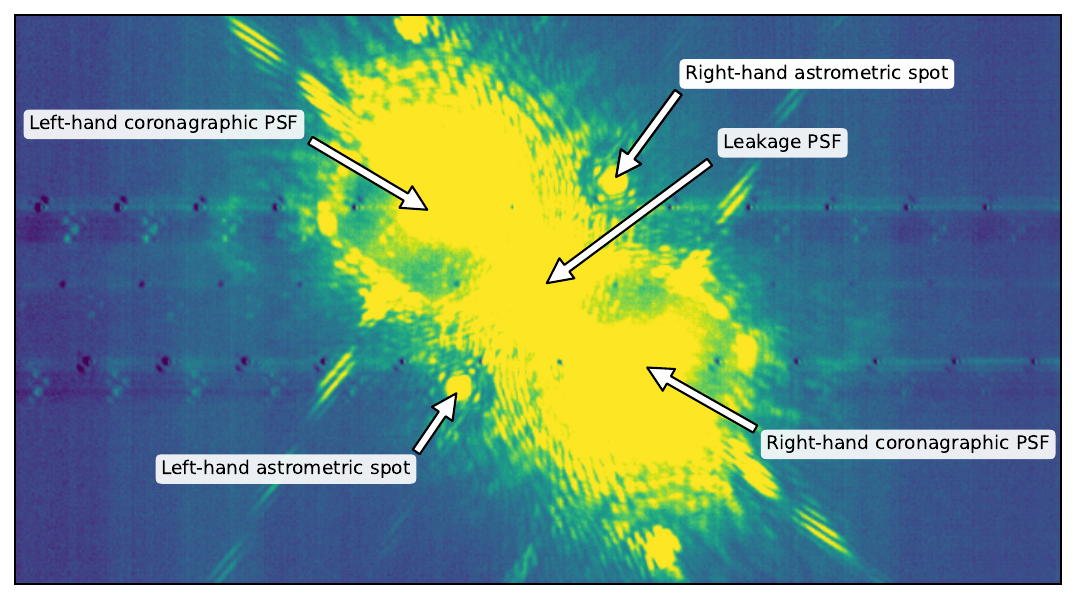}
    \caption{Median combined frame highlighting the impact of the repeating detector noise pattern in the dark holes of the coronagraphic \acp{psf}.
    The orientation of the image is that of the detector; the left- and right-hand notation is used to describe the positions of the respective \acp{psf} relative to this orientation.}
    \label{fig:detector_pattern}
\end{figure*}

Detector artefacts can cause this additional noise component in the dark hole and explain the low sensitivity.
Figure~\ref{fig:detector_pattern} shows a median-combined frame from the $K$-peak observation with a colour scale selected to highlight background variations.
There is a repeating horizontal pattern on the detector correlated with regions of the \ac{psf} with high flux, i.e. the two coronagraphic \acp{psf} and the central leakage term.
Both nod positions used to generate Fig.~\ref{fig:detector_pattern} are affected, as there is both a positive and a negative version of this same pattern.
The negative pattern originates from the nod position shown in this figure, as this is the one that passes through the centres of the side \acp{psf} and the central leakage term.
The positive pattern, which is spatially offset from the negative pattern, aligns with the \ac{psf} at the other nod position (not visible in this figure).
Both the negative and the positive patterns go through both dark holes, leaving little of the dark hole unaffected.
We estimate the effect to be of the order of $10^{-4}$ relative to the peak of the \ac{app} \ac{psf}, with counts of around 4 to 6 compared to $2\times10^{4}$ in the core of the \ac{app} \ac{psf}.
These artefacts also appear in our Br-$\gamma$ observations but not in the Br-$\alpha$ or Br-$\alpha$-cont observations.
We assumed this effect is present in the $L$ band but that the increased thermal background overwhelms this effect.
The reason for this pattern is unclear: neither the Br-$\gamma$ and $K$-peak observations are saturated and both have detector counts of around 20-30k at the centres of the side \acp{psf} and after background subtraction, which corresponds to approximately 50\% of the well depth for the fast readout mode. 
The periodicity of the pattern suggests electronic cross talk between the readout channels of the \ac{eris} detector caused by the large fluxes of the \ac{gvapp} coronagraphic \acp{psf}.
For time series observations, this pattern is likely to introduce significant systematics to the obtained photometry.
However, it may be possible to largely mitigate these systematics by de-trending using the pixel positions of the star at each nod position as de-correlation parameters, since the pattern is directly correlated with the star's location on the detector \citep[e.g.][]{2024MNRAS.531.2168S, 2025MNRAS.544.3191S}.

\subsection{Use cases and recommendations}

Planetary mass limits have been reached with other \ac{gvapp} coronagraphs \citep[e.g.][]{2023AJ....165..216L} and previous \ac{app} coronagraphs of the 180-degree design \citep[e.g.][]{2011ApJ...736L..32Q, 2013ApJ...764....7K, 2015MNRAS.453.2378M, 2015MNRAS.453.2533M}.
If the orbit of a companion is already well constrained, the user can plan observations to ensure that the companion is positioned centrally in one of the \ac{gvapp} dark holes, thereby maximising its signal to noise \citep[e.g.][]{2010ApJ...722L..49Q, 2021MNRAS.506.3224S}.
It may therefore be preferable in some cases to use alternative coronagraphs to initially detect a companion before then using the \ac{eris} \ac{gvapp} for characterisation.

The \ac{eris} \ac{gvapp} design includes features that make it particularly useful for time series studies, such as obtaining measurements of the brightness variability of directly imaged exoplanets and brown dwarf companions.
Photometric and spectrophotometric measurements show that many substellar objects exhibit variability as they rotate, thought to originate from inhomogeneous atmospheric features such as patchy clouds, magnetic spots, aurorae, and non-equilibrium chemistry \cite[e.g.][]{2015ApJ...799..154M, 2019MNRAS.483..480V, 2022ApJ...924...68V, 2020AJ....159..125L, 2024MNRAS.532.2207B, 2024MNRAS.527.6624L, 2025MNRAS.539.3758C, 2025SciA...11v3324T, 2026ApJ...997..136O}.
Measuring the variability of directly imaged companions using ground-based observations is challenging, as a simultaneous photometric reference is required to remove atmospheric and instrumental systematics from the photometry of a companion.
Such a photometric reference is unavailable for most coronagraphic observations, but the leakage term or central astrometric spots of \ac{gvapp} coronagraphs can be used to correct the photometry of a companion positioned in the dark holes when kept unsaturated.
This approach has been previously demonstrated using the central \ac{psf} of the LBT dgvAPP and is described in detail in \citet[][]{2023MNRAS.520.4235S, 2024MNRAS.531.2168S, 2025MNRAS.544.3191S}.

However, we highlight that while the \ac{eris} \ac{gvapp} does provide this built in photometric reference, its throughput and contrast underperformance ($5\times10^{-5}$ on-sky contrast compared to $1\times10^{-5}$ design contrast) may preclude useful variability studies of known exoplanet companions, as it likely will not be possible to detect them with sufficient signal to noise and cadence (see Table~\ref{tab:ppc_sensitivity}).
Variability studies with the \ac{eris} \ac{gvapp} may therefore only be possible for the brightest companions, such as brown dwarfs, or by compromising on cadence to achieve the necessary signal.
Furthermore, we found that there is a slight variance in the brightness of the leakage term relative to the coronagraphic \acp{psf} depending on position on the detector, and the noise patterns shown in Fig.~\ref{fig:detector_pattern} also add position-dependent variance (see Sect.~\ref{sect:imbalances}).
Although both of these effects are at a relatively low level and can be mitigated using de-trending techniques (see Sect.~\ref{sect:reduced_sensitivity}), they will inherently increase the complexity of such observations and may reduce the level of precision that can be achieved. 

In order to conduct time series observations with the \ac{eris} \ac{gvapp}, it is critical that exposure times are chosen such that the simultaneous astrometric and photometric reference \acp{psf} (i.e. the leakage term) remain unsaturated.
We also recommend using the $K$-peak filter for variability studies using the \ac{eris} \ac{gvapp}, as its 6\% width is designed to maximise throughput while limiting wavelength smearing to $<1\lambda/D$ \cite[see][]{2022MNRAS.515.5629D}.

For all \ac{eris} \ac{gvapp} observations, we strongly recommend using an ABBA nodding pattern to facilitate background subtraction while minimising overheads, with a spacing between the two nod positions such that the \ac{gvapp} \acp{psf} do not overlap.
The NIX detector has several bad pixel clusters, which dominate large areas of the frame \citep[e.g. see Fig. 2 in][]{2025A&A...698A..52M}, so the two nod positions should therefore be chosen to avoid these regions. 
This is achieved by setting the alternate nod position to be offset from the default \ac{eris} NIX \ac{gvapp} acquisition position by $-9.75$ arcseconds in the $x$-direction. 
It is recommended to use the \texttt{window4} sub-array, as this frame size remains sufficiently large to fit the \ac{gvapp} \acp{psf} at two nod positions while significantly reducing the volume of the data. 
Further information on the \ac{eris} NIX subarrays and offsetting can be found in the \ac{eris} User Manual\footnote{ERIS User Manual: \url{https://www.eso.org/sci/facilities/paranal/instruments/eris/doc.html}.}.

The shortest integration time in the smallest available detector window limits the use of normal imaging to stars fainter than $m_\text{K} = 2.6$ in the $K$-peak filter and $m_\text{L} = 0.9$ in the Br-$\alpha$-cont filter.
Although brighter stars will saturate, the \ac{eris} \ac{gvapp} leakage term can be still be used as a photometric reference of the target if kept unsaturated, and thus it can still be used for photometrically calibrated observations of brighter targets.

\section{Conclusions}\label{sec:conclusion}

The \ac{gvapp} is built to within the design tolerances specified.
The coronagraph throughput, consisting of the (designed) EE throughput, undersizing of the \ac{app} telescope pupil, transmission due to the liquid crystal and glue layers, and splitting of target flux between the two coronagraphic \acp{psf}, is consistent with other operating \acp{gvapp}.

The phase pattern successfully pushes the stellar diffraction halo below the wind-driven halo resulting from the partial AO correction of atmospheric turbulence.
The on-sky raw contrast is then limited by noise consistent with star light scattered by the optics within the imaging path of \ac{eris}.
$L$-band observations are limited by the large sky background from warm optics in the \ac{eris} optical path and poor transmission due to multiple glue layers in the \ac{eris} \ac{gvapp}.
In testing, the \ac{gvapp} reaches its design specification in transmission with about 90\% in the K bands and 60\% in the $L$ band.
While the \ac{gvapp} reaches a considerable raw contrast laboratory performance of $1\times10^{-5}$, it does not reach the specified $5\times10^{-5}$: electronic detector pattern noise induced by large amounts of stellar flux causes pixel-scale artefacts within the dark holes of the coronagraphic \ac{psf}, limiting the $K$-band sensitivity.

We recommend the \ac{eris} \ac{gvapp} as a tool for studies requiring small \acp{iwa}, as \ac{gvapp} coronagraphs do not use masks to block target stars.
It is particularly suitable for characterising substellar companions with known separations and position angles, which allows them to be positioned in the \ac{gvapp} dark holes for observations \citep[e.g.][]{2010ApJ...722L..49Q, 2021MNRAS.506.3224S, 2022AJ....163..217D}.
Time series observations that use the leakage term as a photometric reference may also be possible for the brightest companions, such as brown dwarfs \citep[e.g.][]{2024MNRAS.531.2168S, 2025MNRAS.544.3191S}; however, we note that the contrast performance likely precludes such studies for known high-contrast exoplanets as it will be challenging to achieve the required signal to noise and cadence.

Improvements for \ac{eris} \ac{gvapp} data can be made by looking into how to minimise or model the detector electronic noise for removal.
Future \ac{gvapp} coronagraphs that use the liquid crystal manufacturing technology can increase the throughput if thinner layers of optically active material and optical glues that do not have strong infrared absorption features in them are used.
Minimising the number of total glue layers in the final assembled optic of future \ac{gvapp} coronagraphs (from \ac{eris}'s four layers down to a minimum of two) will also improve the sensitivity -- these lessons have been taken into consideration with the design of the \acp{gvapp} \citep{2024SPIE13096E..52A} for the \ac{elt} instruments, including METIS \citep{2021Msngr.182...22B}.
METIS is an \ac{elt} instrument with an $L$- and $M$-band spectrometer (R$\sim$100,000) that can be used in combination with \acp{gvapp} designed to work with an image slicer and the imaging channels of the instrument.
With the \ac{elt}'s 39 m diameter, this will allow 2.9-5.3~\textmu m high-resolution spectroscopy and spectral variability measurements to be obtained for directly imaged exoplanets at extremely close separations \citep[e.g.][]{2014Natur.509...63S, 2015A&A...576A..59S, 2023MNRAS.520.4235S, 2025MNRAS.544.3191S, 2024MNRAS.531.2356P, 2025MNRAS.537..134G}.

\section*{Data availability}

 This paper was compiled using the {\tt showyourwork!} framework \citep{Luger2021}, which includes all original data files, the computer scripts used to generate the figures, and the associated compilation framework.
 This is available at \url{https://github.com/mkenworthy/ERIS_gvAPP}
 
\begin{acknowledgements}

We thank our referee for a very careful reading of our manuscript and for the improvements this has made.
JH acknowledges the financial support from the Swiss National Science Foundation (SNSF) under project grant number 200020\_200399.
This work has been carried out within the framework of the NCCR PlanetS supported by the Swiss National Science Foundation under grants 51NF40\_182901 and 51NF40\_205606.
BJS and BAB acknowledge funding by the UK Science and Technology Facilities Council (STFC) grant nos. ST/V000594/1 and UKRI1196.
JLB acknowledges funding from the European Research Council (ERC) under the European Union's Horizon 2020 research and innovation program under grant agreement No 805445.
AZ acknowledges support from ANID -- Millennium Science Initiative Program -- centre Code NCN2024\_001 and Fondecyt Regular grant number 1250249.
This work is based on observations collected at the European Organisation for Astronomical Research in the Southern Hemisphere under ESO programme 112.25QA.001 (PI: Sutlieff).
BJS and XC would like to thank Marco Berton, Robert De Rosa, and the support staff at Paranal Observatory for their assistance in obtaining these observations, and Derek McLeod and Britton Smith for their technical advice on cluster computing.
This work made use of the University of Edinburgh's Institute for Astronomy research computing cluster, {\sl cuillin}, which is partially funded by the UK STFC. We would like to thank Eric Tittley for his work managing and maintaining {\sl cuillin}.
This research has used the SIMBAD database, operated at CDS, Strasbourg, France \citep{wenger2000}.
This work has used data from the European Space Agency (ESA) mission {\it Gaia} (\url{https://www.cosmos.esa.int/gaia}), processed by the {\it Gaia} Data Processing and Analysis Consortium (DPAC, \url{https://www.cosmos.esa.int/web/gaia/dpac/consortium}).
Funding for the DPAC has been provided by national institutions, in particular the institutions participating in the {\it Gaia} Multilateral Agreement.
This research has made use of NASA's Astrophysics Data System.
This research made use of SAOImageDS9, a tool for data visualization supported by the Chandra X-ray Science centre (CXC) and the High Energy Astrophysics Science Archive centre (HEASARC) with support from the JWST Mission office at the Space Telescope Science Institute for 3D visualization \citep{2003ASPC..295..489J}.
To achieve the scientific results presented in this article we made use of the \emph{Python} programming language\footnote{Python Software Foundation, \url{https://www.python.org/}}, especially the \emph{SciPy} \citep{virtanen2020}, \emph{NumPy} \citep{numpy}, \emph{Matplotlib} \citep{Matplotlib}, \emph{emcee} \citep{foreman-mackey2013}, \emph{astropy} \citep{astropy_1,astropy_2,astropy_3}, \emph{scikit-image} \citep{scikit-image}, \emph{Photutils} \citep{larry_bradley_2022_6385735}, \emph{PynPoint} \citep{Amara2012PYNPOINTexoplanets,Stolker2019PynPointdata}, \emph{VIP} \citep{2017AJ....154....7G,2023JOSS....8.4774C}, \emph{TRAP} \citep{Samland2021TRAP:Separations}, and \emph{applefy} \citep{Bonse2023ComparingNoise} packages.

\end{acknowledgements}

\bibliographystyle{aa}
\bibliography{bib}

\begin{thebibliography}{77}
\expandafter\ifx\csname natexlab\endcsname\relax\def\natexlab#1{#1}\fi

\bibitem[{{Absil} {et~al.}(2024){Absil}, {Kenworthy}, {Delacroix}, {Orban de
  Xivry}, {K{\"o}nig}, {Pathak}, {Doelman}, {Por}, {Snik}, {van den Born},
  {Cantalloube}, {Carlotti}, {Courtney-Barrer}, {Forsberg}, {Karlsson},
  {Bertram}, {van Boekel}, {Dolkens}, {Feldt}, {Glauser}, {Pantin}, {Quanz},
  {Bettonvil}, \& {Brandl}}]{2024SPIE13096E..52A}
{Absil}, O., {Kenworthy}, M., {Delacroix}, C., {et~al.} 2024, in Society of
  Photo-Optical Instrumentation Engineers (SPIE) Conference Series, Vol. 13096,
  Ground-based and Airborne Instrumentation for Astronomy X, ed. J.~J.
  {Bryant}, K.~{Motohara}, \& J.~R.~D. {Vernet}, 1309652

\bibitem[{{Amara} \& {Quanz}(2012)}]{Amara2012PYNPOINTexoplanets}
{Amara}, A. \& {Quanz}, S.~P. 2012, \mnras, 427, 948

\bibitem[{{Astropy Collaboration} {et~al.}(2022){Astropy Collaboration},
  {Price-Whelan}, {Lim}, {Earl}, {Starkman}, {Bradley}, {Shupe}, {Patil},
  {Corrales}, {Brasseur}, {N{\"o}the}, {Donath}, {Tollerud}, {Morris},
  {Ginsburg}, {Vaher}, {Weaver}, {Tocknell}, {Jamieson}, {van Kerkwijk},
  {Robitaille}, {Merry}, {Bachetti}, {G{\"u}nther}, {Aldcroft},
  {Alvarado-Montes}, {Archibald}, {B{\'o}di}, {Bapat}, {Barentsen},
  {Baz{\'a}n}, {Biswas}, {Boquien}, {Burke}, {Cara}, {Cara}, {Conroy},
  {Conseil}, {Craig}, {Cross}, {Cruz}, {D'Eugenio}, {Dencheva}, {Devillepoix},
  {Dietrich}, {Eigenbrot}, {Erben}, {Ferreira}, {Foreman-Mackey}, {Fox},
  {Freij}, {Garg}, {Geda}, {Glattly}, {Gondhalekar}, {Gordon}, {Grant},
  {Greenfield}, {Groener}, {Guest}, {Gurovich}, {Handberg}, {Hart},
  {Hatfield-Dodds}, {Homeier}, {Hosseinzadeh}, {Jenness}, {Jones}, {Joseph},
  {Kalmbach}, {Karamehmetoglu}, {Ka{\l}uszy{\'n}ski}, {Kelley}, {Kern},
  {Kerzendorf}, {Koch}, {Kulumani}, {Lee}, {Ly}, {Ma}, {MacBride}, {Maljaars},
  {Muna}, {Murphy}, {Norman}, {O'Steen}, {Oman}, {Pacifici}, {Pascual},
  {Pascual-Granado}, {Patil}, {Perren}, {Pickering}, {Rastogi}, {Roulston},
  {Ryan}, {Rykoff}, {Sabater}, {Sakurikar}, {Salgado}, {Sanghi}, {Saunders},
  {Savchenko}, {Schwardt}, {Seifert-Eckert}, {Shih}, {Jain}, {Shukla}, {Sick},
  {Simpson}, {Singanamalla}, {Singer}, {Singhal}, {Sinha}, {Sip{\H{o}}cz},
  {Spitler}, {Stansby}, {Streicher}, {{\v{S}}umak}, {Swinbank}, {Taranu},
  {Tewary}, {Tremblay}, {de Val-Borro}, {Van Kooten}, {Vasovi{\'c}}, {Verma},
  {de Miranda Cardoso}, {Williams}, {Wilson}, {Winkel}, {Wood-Vasey}, {Xue},
  {Yoachim}, {Zhang}, {Zonca}, \& {Astropy Project Contributors}}]{astropy_3}
{Astropy Collaboration}, {Price-Whelan}, A.~M., {Lim}, P.~L., {et~al.} 2022,
  \apj, 935, 167

\bibitem[{{Astropy Collaboration} {et~al.}(2018){Astropy Collaboration},
  {Price-Whelan}, {Sip{\H o}cz}, {G{\"u}nther}, {Lim}, {Crawford}, {Conseil},
  {Shupe}, {Craig}, {Dencheva}, {Ginsburg}, {VanderPlas}, {Bradley},
  {P{\'e}rez-Su{\'a}rez}, {de Val-Borro}, {Aldcroft}, {Cruz}, {Robitaille},
  {Tollerud}, {Ardelean}, {Babej}, {Bach}, {Bachetti}, {Bakanov}, {Bamford},
  {Barentsen}, {Barmby}, {Baumbach}, {Berry}, {Biscani}, {Boquien}, {Bostroem},
  {Bouma}, {Brammer}, {Bray}, {Breytenbach}, {Buddelmeijer}, {Burke},
  {Calderone}, {Cano Rodr{\'{\i}}guez}, {Cara}, {Cardoso}, {Cheedella},
  {Copin}, {Corrales}, {Crichton}, {D'Avella}, {Deil}, {Depagne}, {Dietrich},
  {Donath}, {Droettboom}, {Earl}, {Erben}, {Fabbro}, {Ferreira}, {Finethy},
  {Fox}, {Garrison}, {Gibbons}, {Goldstein}, {Gommers}, {Greco}, {Greenfield},
  {Groener}, {Grollier}, {Hagen}, {Hirst}, {Homeier}, {Horton}, {Hosseinzadeh},
  {Hu}, {Hunkeler}, {Ivezi{\'c}}, {Jain}, {Jenness}, {Kanarek}, {Kendrew},
  {Kern}, {Kerzendorf}, {Khvalko}, {King}, {Kirkby}, {Kulkarni}, {Kumar},
  {Lee}, {Lenz}, {Littlefair}, {Ma}, {Macleod}, {Mastropietro}, {McCully},
  {Montagnac}, {Morris}, {Mueller}, {Mumford}, {Muna}, {Murphy}, {Nelson},
  {Nguyen}, {Ninan}, {N{\"o}the}, {Ogaz}, {Oh}, {Parejko}, {Parley}, {Pascual},
  {Patil}, {Patil}, {Plunkett}, {Prochaska}, {Rastogi}, {Reddy Janga},
  {Sabater}, {Sakurikar}, {Seifert}, {Sherbert}, {Sherwood-Taylor}, {Shih},
  {Sick}, {Silbiger}, {Singanamalla}, {Singer}, {Sladen}, {Sooley},
  {Sornarajah}, {Streicher}, {Teuben}, {Thomas}, {Tremblay}, {Turner},
  {Terr{\'o}n}, {van Kerkwijk}, {de la Vega}, {Watkins}, {Weaver}, {Whitmore},
  {Woillez}, {Zabalza}, \& {Astropy Contributors}}]{astropy_2}
{Astropy Collaboration}, {Price-Whelan}, A.~M., {Sip{\H o}cz}, B.~M., {et~al.}
  2018, \aj, 156, 123

\bibitem[{{Astropy Collaboration} {et~al.}(2013){Astropy Collaboration},
  {Robitaille}, {Tollerud}, {Greenfield}, {Droettboom}, {Bray}, {Aldcroft},
  {Davis}, {Ginsburg}, {Price-Whelan}, {Kerzendorf}, {Conley}, {Crighton},
  {Barbary}, {Muna}, {Ferguson}, {Grollier}, {Parikh}, {Nair}, {Unther},
  {Deil}, {Woillez}, {Conseil}, {Kramer}, {Turner}, {Singer}, {Fox}, {Weaver},
  {Zabalza}, {Edwards}, {Azalee Bostroem}, {Burke}, {Casey}, {Crawford},
  {Dencheva}, {Ely}, {Jenness}, {Labrie}, {Lim}, {Pierfederici}, {Pontzen},
  {Ptak}, {Refsdal}, {Servillat}, \& {Streicher}}]{astropy_1}
{Astropy Collaboration}, {Robitaille}, T.~P., {Tollerud}, E.~J., {et~al.} 2013,
  \aap, 558, A33

\bibitem[{{Barbato} {et~al.}(2025){Barbato}, {Mesa}, {D'Orazi}, {Desidera},
  {Ruggieri}, {Farinato}, {Marafatto}, {Carolo}, {Vassallo}, {Ertel}, {Hom},
  {Anche}, {Battaini}, {Becker}, {Bergomi}, {Biondi}, {Cardwell}, {Cerpelloni},
  {Chauvin}, {Chinellato}, {Desgrange}, {Di Filippo}, {Dima}, {Machado},
  {Gratton}, {Greggio}, {Henning}, {Kenworthy}, {Laudisio}, {Lazzoni},
  {Leisenring}, {Lessio}, {Lorenzetto}, {Mohr}, {Montoya}, {Rodeghiero},
  {Patience}, {Power}, {Ricci}, {Santhakumari}, {Sozzetti}, {Umbriaco},
  {Pallauta}, {Viotto}, \& {Wagner}}]{2025A&A...693A..81B}
{Barbato}, D., {Mesa}, D., {D'Orazi}, V., {et~al.} 2025, \aap, 693, A81

\bibitem[{Berry(1984)}]{Berry}
Berry, M.~V. 1984, Proceedings of the Royal Society of London A: Mathematical,
  Physical and Engineering Sciences, 392, 45

\bibitem[{{Biller} {et~al.}(2024){Biller}, {Vos}, {Zhou}, {McCarthy}, {Tan},
  {Crossfield}, {Whiteford}, {Suarez}, {Faherty}, {Manjavacas}, {Chen}, {Liu},
  {Sutlieff}, {Limbach}, {Molliere}, {Dupuy}, {Oliveros-Gomez}, {Muirhead},
  {Henning}, {Mace}, {Crouzet}, {Karalidi}, {Morley}, {Tremblin}, \&
  {Kataria}}]{2024MNRAS.532.2207B}
{Biller}, B.~A., {Vos}, J.~M., {Zhou}, Y., {et~al.} 2024, \mnras, 532, 2207

\bibitem[{{Boehle} {et~al.}(2021){Boehle}, {Doelman}, {Konrad}, {Snik},
  {Glauser}, {Por}, {Warriner}, {Shi}, {Escuti}, {Kenworthy}, \&
  {Quanz}}]{2021JATIS...7d5001B}
{Boehle}, A., {Doelman}, D., {Konrad}, B.~S., {et~al.} 2021, Journal of
  Astronomical Telescopes, Instruments, and Systems, 7, 045001

\bibitem[{{Bonse} {et~al.}(2023){Bonse}, {Garvin}, {Gebhard}, {Dannert},
  {Cantalloube}, {Cugno}, {Absil}, {Hayoz}, {Milli}, {Kasper}, \&
  {Quanz}}]{Bonse2023ComparingNoise}
{Bonse}, M.~J., {Garvin}, E.~O., {Gebhard}, T.~D., {et~al.} 2023, \aj, 166, 71

\bibitem[{{Bos} {et~al.}(2019){Bos}, {Doelman}, {Lozi}, {Guyon}, {Keller},
  {Miller}, {Jovanovic}, {Martinache}, \& {Snik}}]{2019A&A...632A..48B}
{Bos}, S.~P., {Doelman}, D.~S., {Lozi}, J., {et~al.} 2019, \aap, 632, A48

\bibitem[{Bradley {et~al.}(2022)Bradley, Sipőcz, Robitaille, Tollerud,
  Vinícius, Deil, Barbary, Wilson, Busko, Donath, Günther, Cara, Lim,
  Meßlinger, Conseil, Bostroem, Droettboom, Bray, Bratholm, Barentsen, Craig,
  Rathi, Pascual, Perren, Georgiev, de~Val-Borro, Kerzendorf, Bach, Quint, \&
  Souchereau}]{larry_bradley_2022_6385735}
Bradley, L., Sipőcz, B., Robitaille, T., {et~al.} 2022, astropy/photutils:

\bibitem[{{Brandl} {et~al.}(2021){Brandl}, {Bettonvil}, {van Boekel},
  {Glauser}, {Quanz}, {Absil}, {Amorim}, {Feldt}, {Glasse}, {G{\"u}del}, {Ho},
  {Labadie}, {Meyer}, {Pantin}, {van Winckel}, \& {METIS
  Consortium}}]{2021Msngr.182...22B}
{Brandl}, B., {Bettonvil}, F., {van Boekel}, R., {et~al.} 2021, The Messenger,
  182, 22

\bibitem[{{Chen} {et~al.}(2025){Chen}, {Biller}, {Tan}, {Vos}, {Zhou},
  {Su{\'a}rez}, {McCarthy}, {Morley}, {Whiteford}, {Dupuy}, {Faherty},
  {Sutlieff}, {Oliveros-Gomez}, {Manjavacas}, {Limbach}, {Lee}, {Karalidi},
  {Crossfield}, {Liu}, {Molliere}, {Muirhead}, {Henning}, {Mace}, {Crouzet}, \&
  {Kataria}}]{2025MNRAS.539.3758C}
{Chen}, X., {Biller}, B.~A., {Tan}, X., {et~al.} 2025, \mnras, 539, 3758

\bibitem[{{Christiaens} {et~al.}(2023){Christiaens}, {Gonzalez}, {Farkas},
  {Dahlqvist}, {Nasedkin}, {Milli}, {Absil}, {Ngo}, {Cantero}, {Rainot},
  {Hammond}, {Bonse}, {Cantalloube}, {Vigan}, {Kompella}, \&
  {Hancock}}]{2023JOSS....8.4774C}
{Christiaens}, V., {Gonzalez}, C., {Farkas}, R., {et~al.} 2023, The Journal of
  Open Source Software, 8, 4774

\bibitem[{{Codona} \& {Angel}(2004)}]{Codona04}
{Codona}, J.~L. \& {Angel}, R. 2004, \apjl, 604, L117

\bibitem[{{Davies} {et~al.}(2023){Davies}, {Absil}, {Agapito}, {Agudo Berbel},
  {Baruffolo}, {Biliotti}, {Black}, {Bonaglia}, {Bonse}, {Briguglio},
  {Campana}, {Cao}, {Carbonaro}, {Cortes}, {Cresci}, {Dallilar}, {Dannert}, {De
  Rosa}, {Deysenroth}, {Di Antonio}, {Di Cianno}, {Di Rico}, {Doelman},
  {Dolci}, {Dorn}, {Eisenhauer}, {Esposito}, {Fantinel}, {Ferruzzi},
  {Feuchtgruber}, {Finger}, {F{\"o}rster Schreiber}, {Gao}, {Gemperlein},
  {Genzel}, {Gillessen}, {Ginski}, {Glauser}, {Glindemann}, {Grani}, {Hartl},
  {Hayoz}, {Heida}, {Henry}, {Hofmann}, {Huber}, {Kasper}, {Keller},
  {Kenworthy}, {Kravchenko}, {Kuntschner}, {Lacour}, {Lightfoot}, {Lunney},
  {Lutz}, {Macintosh}, {Mannucci}, {Marsset}, {Modigliani}, {Neeser}, {Orban de
  Xivry}, {Ott}, {Pallanca}, {Patapis}, {Pearson}, {Pe{\~n}a}, {Percheron},
  {Puglisi}, {Quanz}, {Rabien}, {Rau}, {Riccardi}, {Salasnich}, {Schmid},
  {Schubert}, {Serra}, {Shimizu}, {Snik}, {Sturm}, {Tacconi}, {Taylor},
  {Valentini}, {Waring}, {Wiezorrek}, \& {Xompero}}]{2023A&A...674A.207D}
{Davies}, R., {Absil}, O., {Agapito}, G., {et~al.} 2023, \aap, 674, A207

\bibitem[{{Doelman} {et~al.}(2021){Doelman}, {Snik}, {Por}, {Bos}, {Otten},
  {Kenworthy}, {Haffert}, {Wilby}, {Bohn}, {Sutlieff}, {Miller}, {Ouellet}, {de
  Boer}, {Keller}, {Escuti}, {Shi}, {Warriner}, {Hornburg}, {Birkby}, {Males},
  {Morzinski}, {Close}, {Codona}, {Long}, {Schatz}, {Lumbres}, {Rodack}, {Van
  Gorkom}, {Hedglen}, {Guyon}, {Lozi}, {Groff}, {Chilcote}, {Jovanovic},
  {Thibault}, {de Jonge}, {Allain}, {Vall{\'e}e}, {Patel}, {C{\^o}t{\'e}},
  {Marois}, {Hinz}, {Stone}, {Skemer}, {Briesemeister}, {Boehle}, {Glauser},
  {Taylor}, {Baudoz}, {Huby}, {Absil}, {Carlomagno}, \&
  {Delacroix}}]{2021ApOpt..60D..52D}
{Doelman}, D.~S., {Snik}, F., {Por}, E.~H., {et~al.} 2021, \ao, 60, D52

\bibitem[{{Doelman} {et~al.}(2022){Doelman}, {Stone}, {Briesemeister},
  {Skemer}, {Barman}, {Brock}, {Hinz}, {Bohn}, {Kenworthy}, {Haffert}, {Snik},
  {Ertel}, {Leisenring}, {Woodward}, \& {Skrutskie}}]{2022AJ....163..217D}
{Doelman}, D.~S., {Stone}, J.~M., {Briesemeister}, Z.~W., {et~al.} 2022, \aj,
  163, 217

\bibitem[{{Dubber} {et~al.}(2022){Dubber}, {Biller}, {Bonavita}, {Allers},
  {Fontanive}, {Kenworthy}, {Bonnefoy}, \& {Taylor}}]{2022MNRAS.515.5629D}
{Dubber}, S., {Biller}, B., {Bonavita}, M., {et~al.} 2022, \mnras, 515, 5629

\bibitem[{{Foreman-Mackey} {et~al.}(2013){Foreman-Mackey}, {Hogg}, {Lang}, \&
  {Goodman}}]{foreman-mackey2013}
{Foreman-Mackey}, D., {Hogg}, D.~W., {Lang}, D., \& {Goodman}, J. 2013, \pasp,
  125, 306

\bibitem[{{Gandhi} {et~al.}(2025){Gandhi}, {de Regt}, {Snellen},
  {Palma-Bifani}, {Abdoulwahab}, {Chauvin}, {Gonz{\'a}lez Picos}, {Zhang},
  {Landman}, {Stolker}, {Kesseli}, {Mulder}, {Chomez}, {Lagrange}, \&
  {Zurlo}}]{2025MNRAS.537..134G}
{Gandhi}, S., {de Regt}, S., {Snellen}, I., {et~al.} 2025, \mnras, 537, 134

\bibitem[{{George} {et~al.}(2016){George}, {Gr{\"a}ff}, {Feuchtgruber},
  {Hartl}, {Eisenhauer}, {Buron}, {Davies}, {Genzel}, {Huber}, {Rau},
  {Plattner}, {Wiezorrek}, {Weisz}, {Amico}, {Glindeman}, {Hau}, \&
  {Kuntschner}}]{George16}
{George}, E.~M., {Gr{\"a}ff}, D., {Feuchtgruber}, H., {et~al.} 2016, Proc.
  SPIE, 9908, 99080g

\bibitem[{{Gomez Gonzalez} {et~al.}(2017){Gomez Gonzalez}, {Wertz}, {Absil},
  {Christiaens}, {Defr{\`e}re}, {Mawet}, {Milli}, {Absil}, {Van Droogenbroeck},
  {Cantalloube}, {Hinz}, {Skemer}, {Karlsson}, \&
  {Surdej}}]{2017AJ....154....7G}
{Gomez Gonzalez}, C.~A., {Wertz}, O., {Absil}, O., {et~al.} 2017, \aj, 154, 7

\bibitem[{Harris {et~al.}(2020)Harris, Millman, van~der Walt, Gommers,
  Virtanen, Cournapeau, Wieser, Taylor, Berg, Smith, Kern, Picus, Hoyer, van
  Kerkwijk, Brett, Haldane, del R{\'{i}}o, Wiebe, Peterson,
  G{\'{e}}rard-Marchant, Sheppard, Reddy, Weckesser, Abbasi, Gohlke, \&
  Oliphant}]{numpy}
Harris, C.~R., Millman, K.~J., van~der Walt, S.~J., {et~al.} 2020, Nature, 585,
  357

\bibitem[{{Hunter}(2007)}]{Matplotlib}
{Hunter}, J.~D. 2007, Computing in Science and Engineering, 9, 90

\bibitem[{{Joye} \& {Mandel}(2003)}]{2003ASPC..295..489J}
{Joye}, W.~A. \& {Mandel}, E. 2003, in Astronomical Society of the Pacific
  Conference Series, Vol. 295, Astronomical Data Analysis Software and Systems
  XII, ed. H.~E. {Payne}, R.~I. {Jedrzejewski}, \& R.~N. {Hook}, 489

\bibitem[{{Kenworthy} {et~al.}(2010){Kenworthy}, {Quanz}, {Meyer}, {Kasper},
  {Girard}, {Lenzen}, {Codona}, \& {Hinz}}]{Kenworthy10a}
{Kenworthy}, M., {Quanz}, S., {Meyer}, M., {et~al.} 2010, The Messenger, 141, 2

\bibitem[{{Kenworthy} {et~al.}(2007){Kenworthy}, {Codona}, {Hinz}, {Angel},
  {Heinze}, \& {Sivanandam}}]{Kenworthy07}
{Kenworthy}, M.~A., {Codona}, J.~L., {Hinz}, P.~M., {et~al.} 2007, \apj, 660,
  762

\bibitem[{{Kenworthy} \& {Haffert}(2025)}]{2025ARA&A..63..179K}
{Kenworthy}, M.~A. \& {Haffert}, S.~Y. 2025, \araa, 63, 179

\bibitem[{{Kenworthy} {et~al.}(2013){Kenworthy}, {Meshkat}, {Quanz}, {Girard},
  {Meyer}, \& {Kasper}}]{2013ApJ...764....7K}
{Kenworthy}, M.~A., {Meshkat}, T., {Quanz}, S.~P., {et~al.} 2013, \apj, 764, 7

\bibitem[{Kim {et~al.}(2015)Kim, Li, Miskiewicz, Oh, Kudenov, \&
  Escuti}]{kim2015fabrication}
Kim, J., Li, Y., Miskiewicz, M.~N., {et~al.} 2015, Optica, 2, 958

\bibitem[{{Lenzen} {et~al.}(2003){Lenzen}, {Hartung}, {Brandner}, {Finger},
  {Hubin}, {Lacombe}, {Lagrange}, {Lehnert}, {Moorwood}, \&
  {Mouillet}}]{Lenzen03}
{Lenzen}, R., {Hartung}, M., {Brandner}, W., {et~al.} 2003, in Society of
  Photo-Optical Instrumentation Engineers (SPIE) Conference Series, Vol. 4841,
  Society of Photo-Optical Instrumentation Engineers (SPIE) Conference Series,
  ed. M.~{Iye} \& A.~F.~M. {Moorwood}, 944--952

\bibitem[{{Lew} {et~al.}(2020){Lew}, {Apai}, {Zhou}, {Radigan}, {Marley},
  {Schneider}, {Cowan}, {Miles-P{\'a}ez}, {Manjavacas}, {Karalidi}, {Bedin},
  {Lowrance}, \& {Burgasser}}]{2020AJ....159..125L}
{Lew}, B. W.~P., {Apai}, D., {Zhou}, Y., {et~al.} 2020, \aj, 159, 125

\bibitem[{{Liu} {et~al.}(2024){Liu}, {Biller}, {Vos}, {Whiteford}, {Zhang},
  {Liu}, {Fontanive}, {Manjavacas}, {Henning}, {Kenworthy}, {Bonavita},
  {Bonnefoy}, {Bubb}, {Petrus}, \& {Schlieder}}]{2024MNRAS.527.6624L}
{Liu}, P., {Biller}, B.~A., {Vos}, J.~M., {et~al.} 2024, \mnras, 527, 6624

\bibitem[{{Liu} {et~al.}(2023){Liu}, {Bohn}, {Doelman}, {Sutlieff}, {Samland},
  {Kenworthy}, {Snik}, {Birkby}, {Biller}, {Males}, {Morzinski}, {Close}, \&
  {Otten}}]{LiuP2023}
{Liu}, P., {Bohn}, A.~J., {Doelman}, D.~S., {et~al.} 2023, \aap, 674, A115

\bibitem[{{Long} {et~al.}(2023){Long}, {Males}, {Haffert}, {Pearce}, {Marley},
  {Morzinski}, {Close}, {Otten}, {Snik}, {Kenworthy}, {Keller}, {Hinz},
  {Monnier}, {Weinberger}, \& {Tolls}}]{2023AJ....165..216L}
{Long}, J.~D., {Males}, J.~R., {Haffert}, S.~Y., {et~al.} 2023, \aj, 165, 216

\bibitem[{{Long} {et~al.}(2018){Long}, {Males}, {Morzinski}, {Close}, {Snik},
  {Kenworthy}, {Otten}, {Monnier}, {Tolls}, \&
  {Weinberger}}]{2018SPIE10703E..2TL}
{Long}, J.~D., {Males}, J.~R., {Morzinski}, K.~M., {et~al.} 2018, in Society of
  Photo-Optical Instrumentation Engineers (SPIE) Conference Series, Vol. 10703,
  Adaptive Optics Systems VI, ed. L.~M. {Close}, L.~{Schreiber}, \&
  D.~{Schmidt}, 107032T

\bibitem[{{Luger} {et~al.}(2021){Luger}, {Bedell}, {Foreman-Mackey},
  {Crossfield}, {Zhao}, \& {Hogg}}]{Luger2021}
{Luger}, R., {Bedell}, M., {Foreman-Mackey}, D., {et~al.} 2021, arXiv e-prints,
  arXiv:2110.06271

\bibitem[{{Maio} {et~al.}(2025){Maio}, {Roccatagliata}, {Fedele}, {Garufi},
  {Zurlo}, {Lazzoni}, {Facchini}, {Gratton}, {Mesa}, {Toci}, {Antoniucci},
  {Desidera}, {Pino}, {Rigliaco}, {Codella}, {Podio}, {D'Orazi}, {Lodato},
  {Pedichini}, \& {Testi}}]{2025A&A...698A..52M}
{Maio}, F., {Roccatagliata}, V., {Fedele}, D., {et~al.} 2025, \aap, 698, A52

\bibitem[{{Marois} {et~al.}(2006){Marois}, {Lafreni{\`e}re}, {Doyon},
  {Macintosh}, \& {Nadeau}}]{2006ApJ...641..556M}
{Marois}, C., {Lafreni{\`e}re}, D., {Doyon}, R., {Macintosh}, B., \& {Nadeau},
  D. 2006, \apj, 641, 556

\bibitem[{{Mesa} {et~al.}(2025){Mesa}, {Gratton}, {D'Orazi}, {Carolo},
  {Vassallo}, {Farinato}, {Marafatto}, {Wagner}, {Hom}, {Ertel}, {Henning},
  {Desgrange}, {Barbato}, {Bergomi}, {Cerpelloni}, {Desidera}, {Di Filippo},
  {Doelman}, {Machado}, {Greggio}, {Grenz}, {Kenworthy}, {Laudisio}, {Lazzoni},
  {Leisenring}, {Lorenzetto}, {Santhakumari}, {Ricci}, {Snik}, {Umbriaco},
  {Pallauta}, \& {Viotto}}]{2025MNRAS.536.1455M}
{Mesa}, D., {Gratton}, R., {D'Orazi}, V., {et~al.} 2025, \mnras, 536, 1455

\bibitem[{{Meshkat} {et~al.}(2015{\natexlab{a}}){Meshkat}, {Bonnefoy},
  {Mamajek}, {Quanz}, {Chauvin}, {Kenworthy}, {Rameau}, {Meyer}, {Lagrange},
  {Lannier}, \& {Delorme}}]{2015MNRAS.453.2378M}
{Meshkat}, T., {Bonnefoy}, M., {Mamajek}, E.~E., {et~al.} 2015{\natexlab{a}},
  \mnras, 453, 2378

\bibitem[{{Meshkat} {et~al.}(2014){Meshkat}, {Kenworthy}, {Quanz}, \&
  {Amara}}]{Meshkat14}
{Meshkat}, T., {Kenworthy}, M.~A., {Quanz}, S.~P., \& {Amara}, A. 2014, \apj,
  780, 17

\bibitem[{{Meshkat} {et~al.}(2015{\natexlab{b}}){Meshkat}, {Kenworthy},
  {Reggiani}, {Quanz}, {Mamajek}, \& {Meyer}}]{2015MNRAS.453.2533M}
{Meshkat}, T., {Kenworthy}, M.~A., {Reggiani}, M., {et~al.} 2015{\natexlab{b}},
  \mnras, 453, 2533

\bibitem[{{Metchev} {et~al.}(2015){Metchev}, {Heinze}, {Apai}, {Flateau},
  {Radigan}, {Burgasser}, {Marley}, {Artigau}, {Plavchan}, \&
  {Goldman}}]{2015ApJ...799..154M}
{Metchev}, S.~A., {Heinze}, A., {Apai}, D., {et~al.} 2015, \apj, 799, 154

\bibitem[{{Miller} {et~al.}(2019){Miller}, {Males}, {Guyon}, {Close},
  {Doelman}, {Snik}, {Por}, {Wilby}, {Keller}, {Bohlman}, {Van Gorkom},
  {Rodack}, {Knight}, {Lumbres}, {Bos}, \& {Jovanovic}}]{2019JATIS...5d9004M}
{Miller}, K., {Males}, J.~R., {Guyon}, O., {et~al.} 2019, Journal of
  Astronomical Telescopes, Instruments, and Systems, 5, 049004

\bibitem[{{Miller} {et~al.}(2021){Miller}, {Bos}, {Lozi}, {Guyon}, {Doelman},
  {Vievard}, {Sahoo}, {Deo}, {Jovanovic}, {Martinache}, {Snik}, \&
  {Currie}}]{2021A&A...646A.145M}
{Miller}, K.~L., {Bos}, S.~P., {Lozi}, J., {et~al.} 2021, \aap, 646, A145

\bibitem[{{Oliveros-Gomez} {et~al.}(2026){Oliveros-Gomez}, {Manjavacas},
  {Karalidi}, {Phillippe}, {Campos Estrada}, {Biller}, {Vos}, {Faherty},
  {Chen}, {Dupuy}, {Henning}, {McCarthy}, {Muirhead}, {Lee}, {Tremblin},
  {Ramirez}, {Suarez}, {Sutlieff}, {Tan}, \& {Crouzet}}]{2026ApJ...997..136O}
{Oliveros-Gomez}, N., {Manjavacas}, E., {Karalidi}, T., {et~al.} 2026, \apj,
  997, 136

\bibitem[{{Orban de Xivry} {et~al.}(2024){Orban de Xivry}, {Absil}, {De Rosa},
  {Bonse}, {Dannert}, {Hayoz}, {Grani}, {Puglisi}, {Baruffolo}, {Salasnich},
  {Davies}, {Glauser}, {Huby}, {Kenworthy}, {Quanz}, {Taylor}, \&
  {Zins}}]{2024SPIE13097E..15O}
{Orban de Xivry}, G., {Absil}, O., {De Rosa}, R.~J., {et~al.} 2024, in Society
  of Photo-Optical Instrumentation Engineers (SPIE) Conference Series, Vol.
  13097, Adaptive Optics Systems IX, ed. K.~J. {Jackson}, D.~{Schmidt}, \&
  E.~{Vernet}, 1309715

\bibitem[{Otten {et~al.}(2017)Otten, Snik, Kenworthy, Keller, Males, Morzinski,
  Close, Codona, Hinz, Hornburg, {et~al.}}]{otten2017sky}
Otten, G.~P., Snik, F., Kenworthy, M.~A., {et~al.} 2017, The Astrophysical
  Journal, 834, 175

\bibitem[{{Otten} {et~al.}(2014){Otten}, {Snik}, {Kenworthy}, {Miskiewicz}, \&
  {Escuti}}]{Otten14}
{Otten}, G.~P.~P.~L., {Snik}, F., {Kenworthy}, M.~A., {Miskiewicz}, M.~N., \&
  {Escuti}, M.~J. 2014, Optics Express, 22, 30287

\bibitem[{Pancharatnam(1956)}]{Pancharatnam}
Pancharatnam, S. 1956, Proceedings of the Indian Academy of Sciences, Section
  A, 44, 247

\bibitem[{{Parker} {et~al.}(2024){Parker}, {Birkby}, {Landman}, {Wardenier},
  {Young}, {Vaughan}, {van Sluijs}, {Brogi}, {Parmentier}, \&
  {Line}}]{2024MNRAS.531.2356P}
{Parker}, L.~T., {Birkby}, J.~L., {Landman}, R., {et~al.} 2024, \mnras, 531,
  2356

\bibitem[{{Pearson} {et~al.}(2016){Pearson}, {Taylor}, {Davies}, {MacIntosh},
  {Henry}, {Lunney}, {Waring}, {Gao}, {Lightfoot}, {Glauser}, {Quanz}, {Meyer},
  {Schmid}, {March}, {Bachmann}, {Feuchtgruber}, {George}, {Sturm}, {Biller},
  {Hinckley}, {Kenworthy}, {Amico}, {Glindemann}, {Kasper}, {Kuntschner},
  {Dorn}, \& {Egner}}]{2016SPIE.9908E..3FP}
{Pearson}, D., {Taylor}, W., {Davies}, R., {et~al.} 2016, in Society of
  Photo-Optical Instrumentation Engineers (SPIE) Conference Series, Vol. 9908,
  Ground-based and Airborne Instrumentation for Astronomy VI, ed. C.~J.
  {Evans}, L.~{Simard}, \& H.~{Takami}, 99083F

\bibitem[{Por {et~al.}(2018)Por, Haffert, Radhakrishnan, Doelman, Van~Kooten,
  \& Bos}]{por2018hcipy}
Por, E.~H., Haffert, S.~Y., Radhakrishnan, V.~M., {et~al.} 2018, in Proc.
  {{SPIE}}, Vol. 10703, Adaptive Optics Systems VI

\bibitem[{{Quanz} {et~al.}(2013){Quanz}, {Amara}, {Meyer}, {Kenworthy},
  {Kasper}, \& {Girard}}]{Quanz13}
{Quanz}, S.~P., {Amara}, A., {Meyer}, M.~R., {et~al.} 2013, \apjl, 766, L1

\bibitem[{{Quanz} {et~al.}(2011){Quanz}, {Kenworthy}, {Meyer}, {Girard}, \&
  {Kasper}}]{2011ApJ...736L..32Q}
{Quanz}, S.~P., {Kenworthy}, M.~A., {Meyer}, M.~R., {Girard}, J. H.~V., \&
  {Kasper}, M. 2011, \apjl, 736, L32

\bibitem[{{Quanz} {et~al.}(2010){Quanz}, {Meyer}, {Kenworthy}, {Girard},
  {Kasper}, {Lagrange}, {Apai}, {Boccaletti}, {Bonnefoy}, {Chauvin}, {Hinz}, \&
  {Lenzen}}]{2010ApJ...722L..49Q}
{Quanz}, S.~P., {Meyer}, M.~R., {Kenworthy}, M.~A., {et~al.} 2010, \apjl, 722,
  L49

\bibitem[{{Rousset} {et~al.}(2003){Rousset}, {Lacombe}, {Puget}, {Hubin},
  {Gendron}, {Fusco}, {Arsenault}, {Charton}, {Feautrier}, {Gigan}, {Kern},
  {Lagrange}, {Madec}, {Mouillet}, {Rabaud}, {Rabou}, {Stadler}, \&
  {Zins}}]{Rousset03}
{Rousset}, G., {Lacombe}, F., {Puget}, P., {et~al.} 2003, in Society of
  Photo-Optical Instrumentation Engineers (SPIE) Conference Series, Vol. 4839,
  Society of Photo-Optical Instrumentation Engineers (SPIE) Conference Series,
  ed. P.~L. {Wizinowich} \& D.~{Bonaccini}, 140--149

\bibitem[{{Ruane} {et~al.}(2018){Ruane}, {Riggs}, {Mazoyer}, {Por}, {N'Diaye},
  {Huby}, {Baudoz}, {Galicher}, {Douglas}, {Knight}, {Carlomagno}, {Fogarty},
  {Pueyo}, {Zimmerman}, {Absil}, {Beaulieu}, {Cady}, {Carlotti}, {Doelman},
  {Guyon}, {Haffert}, {Jewell}, {Jovanovic}, {Keller}, {Kenworthy}, {Kuhn},
  {Miller}, {Sirbu}, {Snik}, {Wallace}, {Wilby}, \&
  {Ygouf}}]{2018SPIE10698E..2SR}
{Ruane}, G., {Riggs}, A., {Mazoyer}, J., {et~al.} 2018, in Society of
  Photo-Optical Instrumentation Engineers (SPIE) Conference Series, Vol. 10698,
  Space Telescopes and Instrumentation 2018: Optical, Infrared, and Millimeter
  Wave, ed. M.~{Lystrup}, H.~A. {MacEwen}, G.~G. {Fazio}, N.~{Batalha},
  N.~{Siegler}, \& E.~C. {Tong}, 106982S

\bibitem[{{Samland} {et~al.}(2021){Samland}, {Bouwman}, {Hogg}, {Brandner},
  {Henning}, \& {Janson}}]{Samland2021TRAP:Separations}
{Samland}, M., {Bouwman}, J., {Hogg}, D.~W., {et~al.} 2021, \aap, 646, A24

\bibitem[{{Snellen} {et~al.}(2015){Snellen}, {de Kok}, {Birkby}, {Brandl},
  {Brogi}, {Keller}, {Kenworthy}, {Schwarz}, \& {Stuik}}]{2015A&A...576A..59S}
{Snellen}, I., {de Kok}, R., {Birkby}, J.~L., {et~al.} 2015, \aap, 576, A59

\bibitem[{{Snellen} {et~al.}(2014){Snellen}, {Brandl}, {de Kok}, {Brogi},
  {Birkby}, \& {Schwarz}}]{2014Natur.509...63S}
{Snellen}, I. A.~G., {Brandl}, B.~R., {de Kok}, R.~J., {et~al.} 2014, \nat,
  509, 63

\bibitem[{{Snik} {et~al.}(2012){Snik}, {Otten}, {Kenworthy}, {Miskiewicz},
  {Escuti}, {Packham}, \& {Codona}}]{Snik12}
{Snik}, F., {Otten}, G., {Kenworthy}, M., {et~al.} 2012, in Society of
  Photo-Optical Instrumentation Engineers (SPIE) Conference Series, Vol. 8450,
  Society of Photo-Optical Instrumentation Engineers (SPIE) Conference Series

\bibitem[{{Stolker} {et~al.}(2019){Stolker}, {Bonse}, {Quanz}, {Amara},
  {Cugno}, {Bohn}, \& {Boehle}}]{Stolker2019PynPointdata}
{Stolker}, T., {Bonse}, M.~J., {Quanz}, S.~P., {et~al.} 2019, \aap, 621, A59

\bibitem[{{Sutlieff} {et~al.}(2024){Sutlieff}, {Birkby}, {Stone}, {Derkink},
  {Backs}, {Doelman}, {Kenworthy}, {Bohn}, {Ertel}, {Snik}, {Woodward},
  {Ilyin}, {Skemer}, {Leisenring}, {Strassmeier}, {Wang}, {Charbonneau}, \&
  {Biller}}]{2024MNRAS.531.2168S}
{Sutlieff}, B.~J., {Birkby}, J.~L., {Stone}, J.~M., {et~al.} 2024, \mnras, 531,
  2168

\bibitem[{{Sutlieff} {et~al.}(2023){Sutlieff}, {Birkby}, {Stone}, {Doelman},
  {Kenworthy}, {Panwar}, {Bohn}, {Ertel}, {Snik}, {Woodward}, {Skemer},
  {Leisenring}, {Strassmeier}, \& {Charbonneau}}]{2023MNRAS.520.4235S}
{Sutlieff}, B.~J., {Birkby}, J.~L., {Stone}, J.~M., {et~al.} 2023, \mnras, 520,
  4235

\bibitem[{{Sutlieff} {et~al.}(2021){Sutlieff}, {Bohn}, {Birkby}, {Kenworthy},
  {Morzinski}, {Doelman}, {Males}, {Snik}, {Close}, {Hinz}, \&
  {Charbonneau}}]{2021MNRAS.506.3224S}
{Sutlieff}, B.~J., {Bohn}, A.~J., {Birkby}, J.~L., {et~al.} 2021, \mnras, 506,
  3224

\bibitem[{{Sutlieff} {et~al.}(2025){Sutlieff}, {Doelman}, {Birkby},
  {Kenworthy}, {Stone}, {Snik}, {Ertel}, {Biller}, {Woodward}, {Skemer},
  {Leisenring}, {Bohn}, \& {Parker}}]{2025MNRAS.544.3191S}
{Sutlieff}, B.~J., {Doelman}, D.~S., {Birkby}, J.~L., {et~al.} 2025, \mnras,
  544, 3191

\bibitem[{{Tan} {et~al.}(2025){Tan}, {Zhang}, {Marley}, {Zhou}, {Lew}, {Miles},
  {Batalha}, {Biller}, {Chauvin}, {Hinkley}, {Hoch}, {Manjavacas}, {Metchev},
  {Petrus}, {Rickman}, {Skemer}, {Su{\'a}rez}, {Sutlieff}, {Vos}, \&
  {Whiteford}}]{2025SciA...11v3324T}
{Tan}, X., {Zhang}, X., {Marley}, M.~S., {et~al.} 2025, Science Advances, 11,
  22.3324

\bibitem[{van~der Walt {et~al.}(2014)van~der Walt, {S}ch\"onberger,
  {Nunez-Iglesias}, {B}oulogne, {W}arner, {Y}ager, {G}ouillart, {Y}u, \& the
  scikit-image contributors}]{scikit-image}
van~der Walt, S., {S}ch\"onberger, J.~L., {Nunez-Iglesias}, J., {et~al.} 2014,
  PeerJ, 2, e453

\bibitem[{{Virtanen} {et~al.}(2020){Virtanen}, {Gommers}, {Oliphant},
  {Haberland}, {Reddy}, {Cournapeau}, {Burovski}, {Peterson}, {Weckesser},
  {Bright}, {van der Walt}, {Brett}, {Wilson}, {Millman}, {Mayorov}, {Nelson},
  {Jones}, {Kern}, {Larson}, {Carey}, {Polat}, {Feng}, {Moore}, {Vand erPlas},
  {Laxalde}, {Perktold}, {Cimrman}, {Henriksen}, {Quintero}, {Harris},
  {Archibald}, {Ribeiro}, {Pedregosa}, {van Mulbregt}, \& {SciPy 1. 0
  Contributors}}]{virtanen2020}
{Virtanen}, P., {Gommers}, R., {Oliphant}, T.~E., {et~al.} 2020, Nature
  Methods, 17, 261

\bibitem[{{Vos} {et~al.}(2019){Vos}, {Biller}, {Bonavita}, {Eriksson}, {Liu},
  {Best}, {Metchev}, {Radigan}, {Allers}, {Janson}, {Buenzli}, {Dupuy},
  {Bonnefoy}, {Manjavacas}, {Brandner}, {Crossfield}, {Deacon}, {Henning},
  {Homeier}, {Kopytova}, \& {Schlieder}}]{2019MNRAS.483..480V}
{Vos}, J.~M., {Biller}, B.~A., {Bonavita}, M., {et~al.} 2019, \mnras, 483, 480

\bibitem[{{Vos} {et~al.}(2022){Vos}, {Faherty}, {Gagn{\'e}}, {Marley},
  {Metchev}, {Gizis}, {Rice}, \& {Cruz}}]{2022ApJ...924...68V}
{Vos}, J.~M., {Faherty}, J.~K., {Gagn{\'e}}, J., {et~al.} 2022, \apj, 924, 68

\bibitem[{{Wagner} {et~al.}(2020){Wagner}, {Stone}, {Dong}, {Ertel}, {Apai},
  {Doelman}, {Bohn}, {Najita}, {Brittain}, {Kenworthy}, {Keppler}, {Webster},
  {Mailhot}, \& {Snik}}]{2020AJ....159..252W}
{Wagner}, K., {Stone}, J., {Dong}, R., {et~al.} 2020, \aj, 159, 252

\bibitem[{{Wenger} {et~al.}(2000){Wenger}, {Ochsenbein}, {Egret}, {Dubois},
  {Bonnarel}, {Borde}, {Genova}, {Jasniewicz}, {Lalo{\"e}}, {Lesteven}, \&
  {Monier}}]{wenger2000}
{Wenger}, M., {Ochsenbein}, F., {Egret}, D., {et~al.} 2000, \aaps, 143, 9

\end{thebibliography}

\end{document}